\documentclass[traditabstract]{aa} 
\usepackage{graphicx}
\usepackage[varg]{txfonts}
\usepackage{natbib}
\bibpunct{(}{)}{;}{a}{}{,}
\usepackage{array}
\usepackage{xspace}
\usepackage{lscape}
\usepackage{url}
\usepackage{arydshln}
\usepackage{amsmath}
\usepackage{color}
\usepackage[bottom]{footmisc}
\usepackage[hyperfootnotes=false, linktocpage=true, breaklinks=true, colorlinks=true, linkcolor=blue, citecolor=blue, urlcolor=blue]{hyperref}
\usepackage[all]{hypcap}
\usepackage{amsfonts}
\usepackage{amssymb}

\newcommand{\slab}{{\tt slab}\xspace}

\newcommand{\amol}{{\tt amol}\xspace}
\newcommand{\hot}{{\tt hot}\xspace}
\newcommand{\pow}{{\tt pow}\xspace}

\newcommand{\user}{{\tt musr}\xspace}
\newcommand{\xmm}{{\rm XMM$-$\it Newton}\xspace}
\newcommand{\chandra}{{\it Chandra}\xspace}
\newcommand{\spex}{{SPEX}\xspace}
\newcommand{\xstar}{{XSTAR}\xspace}

\newcommand{\oi}{\ion{O}{i}\xspace}
\newcommand{\oii}{\ion{O}{ii}\xspace}
\newcommand{\oiii}{\ion{O}{iii}\xspace}
\newcommand{\ovi}{\ion{O}{vi}\xspace}
\newcommand{\ov}{\ion{O}{v}\xspace}
\newcommand{\ovii}{\ion{O}{vii}\xspace}
\newcommand{\oviii}{\ion{O}{viii}\xspace}
\newcommand{\oiv}{\ion{O}{iv}\xspace}

\mathchardef\mhyphen="2D

\mathchardef\mhyphen="2D

\begin{document}

\title{Interstellar oxygen along the line of sight of Cygnus X-2}

\author{
I. Psaradaki \inst{1,2}
\and
E. Costantini \inst{1,2}
\and 
M. Mehdipour \inst{1}
\and 
D. Rogantini \inst{1,2}
\and 
C. P. de Vries  \inst{1}
\and
F. de Groot \inst{3}
\and 
H. Mutschke \inst{4}
\and
S. Trasobares \inst{5}
\and
L.B.F.M. Waters \inst{1,2}
\and
S.T. Zeegers \inst{6}
}

\institute{
SRON Netherlands Institute for Space Research, Sorbonnelaan 2, 3584 CA Utrecht, the Netherlands\\ \email{I.Psaradaki@sron.nl}
\and
Anton Pannekoek Astronomical Institute, University of Amsterdam, P.O. Box 94249, 1090 GE Amsterdam, the Netherlands
\and 
Debye Institute for Nanomaterials Science, Utrecht University, Universiteitsweg 99, 3584 CG Utrecht, Netherlands
\and 
Astrophysikalisches Institut und Universitats-Sternw\"arte (AIU), Schillerg\"a{\ss}chen 2-3, 07745 Jena, Germany
\and
Ciencia de los Materiales e Ingenier\'ia Metal\'urgica y Qu\'imica Inorg\'anica, Universidad de Cadiz
\and 
Academia Sinica, Institute of Astronomy and Astrophysics, 11F Astronomy-Mathematics Building, NTU/AS campus, No. 1, Section 4, Roosevelt Rd., Taipei 10617, Taiwan}
%


\abstract
{Interstellar dust permeates our Galaxy and plays an important role in many physical processes in the diffuse and dense regions of the interstellar medium. High-resolution X-ray spectroscopy, coupled with modelling based on laboratory dust measurements, provides a unique probe to investigate the interstellar dust properties along our line of sight towards Galactic X-ray sources. Here, we focus on the oxygen content of the interstellar medium through its absorption features in the X-ray spectra. To model the dust features, we perform a laboratory experiment using the electron microscope facility located at the University of Cadiz in Spain, where we acquire new laboratory data in the oxygen K-edge. We study 18 dust samples of silicates and oxides with different chemical compositions. The laboratory measurements are adopted for our astronomical data analysis. We carry out a case study on the X-ray spectrum of the bright low-mass X-ray binary Cygnus X-2, observed by \xmm. We determine different temperature phases of the ISM, and parameterize oxygen in both gas (neutral and ionised) and dust form. We find Solar abundances of oxygen along the line of sight towards the source. Due to both the relatively low depletion of oxygen into dust form and the shape of the oxygen cross section profiles, it is challenging to determine the precise chemistry of interstellar dust. However, silicates provide an acceptable fit. Finally, we discuss the systematic discrepancies in the atomic (gaseous phase) data of the oxygen edge spectral region using different X-ray atomic databases, and also consider future prospects for studying the ISM with the Arcus concept mission.}
\keywords{astrochemistry, X-rays: binaries, ISM:dust}
\authorrunning{I. Psaradaki et al.}
\titlerunning{Interstellar oxygen along the line of sight of Cygnus X-2}
\maketitle

\section{Introduction}
The composition of the interstellar medium (ISM) is very important for the evolution of the Galaxy and for star formation processes. In particular, interstellar dust (ID) is an important constituent of our Galaxy as it can control the temperature of the ISM and it is the catalyst for the formation of complex molecules (\citealt{Mathis1990}).  \\
 The dust chemical and physical properties in the diffuse ISM are not yet fully understood. Observations show that ID in the diffuse ISM is not spacially homogeneous (e.g. \citealt{Planck1}, \citealt{Planck2}, \citealt{Ysard2015}). Silicate grains are an important and abundant component of ID and can be found in many different stages of the life cycle of stars \citep{Henning2010}. It is believed to be mainly produced in oxygen-rich asymptotic giant branch stars \citep{Gail2009} but also in novae and supernovae type II (\citealt{Wooden1993}, \citealt{Rho2008}) and young stellar objects \citep{Nittler1997}. On the other hand, carbonaceous grains could form in interstellar clouds, as carbon grains likely have shorter lifetimes \citep{Jones2011}.\\
 The physical and chemical properties of silicate dust have been historically studied along different directions in the Galaxy at wavelengths ranging from radio to the ultraviolet (\citealt{Draine2001}, \citealt{Dwek2004}). It is believed that the elements such as O, Fe, Si and Mg form the cosmic silicates, which represent most of the dust mass in the ISM \citep{Mathis1998}. Another evidence for such elements to be locked up in dust is given by the fact that they are depleted from the gaseous phase (\citealt{Henning2010}, \citealt{Savage1996}, \citealt{Jenkins2009}). \\
 Oxygen is one of the most abundant and important elements for life on Earth. However, the total budget of oxygen in the diffuse and dense interstellar media cannot be fully explained yet \citep{Whittet2010}. \citet{Jenkins2009} reported that the oxygen in the diffuse interstellar gas is being depleted at a rate that cannot be associated with only silicates and metallic oxides. They found that the total fraction of the missing oxygen is not compatible with the stoichiometric ratios of even the most oxygen rich silicate compounds. This requires some contribution of additional compounds. \\
It has also been proposed that oxygen must be locked up in molecules, together with elements with a high cosmic abundance such as CO, $\rm CO_{2}$ and $ \rm O_{2}$ \citep{Jenkins2009}. These oxygen-bearing molecules have small but not negligible abundance \citep{Ewine2004}. Also, \citet{Ewine1998} found that close to young stellar objects, for large extinction values, oxygen atoms are found in the form of water. \citet{Whittet1988}, \citet{Eiroa1989} and \citet{Smith1993} found that the strength of the 3.05 $\mu$m water ice feature becomes much weaker as the extinction decreases. Later on, \citet{Whittet2001} showed that the same feature becomes detectable for extinction larger than 3.2 and increases linearly with increasing extinction above that value.  \\
High-resolution X-ray spectroscopy is a powerful tool to unveil the physical and chemical properties of the diffuse interstellar medium (\citealt{Wilms}, \citealt{Lee2005}, \citealt{Lee2009}, \citealt{devries2009}, \citealt{Pinto2013}, \citealt{Corrales2016}, \citealt{Schulz}, \citealt{Zeegers2017,Zeegers2019}, \citealt{Rogantini2018}). It provides the possibility to study the composition of cosmic dust in different environments of the ISM through the study of absorption features of gas and dust. In particular, the X-ray Absorption Fine Structures (XAFS) are unique fingerprints of dust. When an X-ray photon excites a core electron (n=1), the outwardgoing photo-electron will interact with the neighboring atoms. This interaction will modify the wave function of the photo-electron due to constructive and destructive interferences. In this way, the absorption probability will be modified in a unique pattern which depends on the configuration of the neighboring atoms. Therefore, the shape of XAFS can be used to determine the chemical structure of dust (see also \citet{Newville} for further explanation).\\
The oxygen K-edge region has been previously explored using observations from the \xmm and the $\chandra$ satellites. \citet{Costantini2012}, \citet{Pinto2010} found that about 15-25 $\%$ of the total neutral oxygen is in dust. On the other hand, \citet{Gatuzz2014} suggested that the oxygen edge can be fitted only with gaseous oxygen. Later, \citet{Eckersall2017} studied the gas features in the oxygen K-edge region with the models used in \citet{Gatuzz2015}. They noticed strong residuals in the fit and they mentioned that they are likely due to dust. Furthermore, \citet{Joachimi2016} searched for evidence of CO in \xmm spectra of low-mass X-ray binaries and suggested two weak detections.\\
In the literature there are different studies regarding the atomic data in the oxygen K-edge region. Theoretical calculations on the K-shell photoabsorption of oxygen ions have been performed by \citet{Garcia2005} and \citet{McLaughlin} using the R-matrix approach. Moreover, laboratory measurements have been carried out in order to determine the absolute energy of the photoabsorption cross sections of oxygen ions (\citealt{Menzel}, \citealt{Stolte}). \citet{Gorczyca} present an analytical formula to describe with accuracy the photoabsorption cross section of $\oi$ in X-ray spectral modelling and \citet{Leutenegger} discuss an experimental technique for the determination of the absolute calibration of the oxygen $\rm K_{\alpha}$ transition energy. Finally, \citet{Frati} review the oxygen K-edge X-ray absorption spectra of atoms, molecules and solids, where the oxygen 1s core electron is excited to the lowest empty states at $\sim 530$ eV.  \\
In order to determine the presence of dust in the oxygen K-edge, we need to use accurate dust models. To this end, we perform a laboratory measuring campaign to compute the absorption cross sections of 18 dust compounds with different chemical composition, such as different types of silicates and oxides. In order to investigate the crystallinity, our samples contain both crystalline and amorphous dust grains. From studies of the 10 $\mu$m silicate feature in the mid-IR band it has been found that the stoichiometry of silicate dust is mostly amorphous olivines ($\rm Mg_{2-x}Fe_{x}SiO_{4}$) and pyroxenes ($\rm Mg_{1-x}Fe_{x}SiO_{3}$) (\citealt{Kemper2004},  \citealt{Min2007}). Also, \citet{Kemper2004}  found that towards SgrA* most silicates have an amorphous structure and less than 2.2$\%$ are crystalline. They have also reported that $\sim$ 85 $\%$ of the amorphous grains are olivines and $\sim$15 $\%$ pyroxenes. The pyroxenes are found to be slightly Mg-rich with Mg/(Mg+Fe) $\sim$ 0.55, while the olivines may be Fe-rich. \citet{Min2007} conclude that the amorphous silicates are Mg-rich with Mg/(Mg+Fe) $\sim$ 0.9. They also find that the crystallinity in the ISM is small and they report a pure crystalline forsterite grain composition ($\rm Mg_{2}SiO_{4}$). Moreover, experimental studies of astrophysical dust analogues with application to infrared observations have been presented by \citet{Nuth2}. \\
We apply the new dust models to the high-resolution \xmm RGS spectrum of Cygnus X-2 in order to study the oxygen K-edge spectral region. It is a bright X-ray source with a moderate column density ( $\rm \sim 2 \times10^{21} cm^{-2}$,  \citealt{Juett2004}) and high flux ($\rm 2.3 \times 10^{-9} erg/sec/cm^{2}$ in the 0.3-2 keV band). Cygnus X-2 has been proposed as a Z-track source (\citealt{Hasinger1989}) and believed to reach the Eddington luminosity limit during its high state (\citealt{Church2010}, \citealt{Mondal2018}, \citealt{King1999}). The distance of the source has been estimated to be around 7 $-$ 12 kpc and its galactic coordinates are $l, b = 87^{o}.33 $-$11^{o}.32$, which means that it is located about 1.4 - 2.4 kpc away from the Galactic plane (\citealt{Cowley}, \citealt{McClintock}, \citealt{Smale}, \citealt{Yao2009}). The systemic velocity of the binary system was found to be about -220 km/s (\citealt{Cowley}, \citealt{Casares}).\\ Previous studies of Cygnus X-2 with Chandra gratings showed that the oxygen region is possibly mildly absorbed by interstellar dust (\citealt{Yao2009}). Moreover, \citet{Juett2004} studied the $\chandra$ HETGS spectrum of Cygnus X-2 and they mentioned that features from dust or molecular compounds are not essential to obtain a good fit. The goal of this investigation is to understand the contribution of dust to the oxygen K-edge spectral region using up-to-date laboratory and atomic data measurements. \\
 The paper is organised as follows. In Section \ref{lab} we describe the dust samples used in this work, the laboratory experiment and the treatment of the laboratory data. In Section \ref{labtocs} and Appendix \ref{appendix} we present the methodology used to calculate the dust extinction cross sections for the oxygen K-edge. In Section \ref{astodata} we apply the new models to the RGS spectrum of Cygnus X-2. The discussion of our results is presented in Section \ref{discussion}, and in Section \ref{atomic} we compare the results obtained using both the atomic databases of \spex and \xstar for the \oi, \oii and \oiii lines. In Section \ref{arcusss} we discuss the prospects of observing the oxygen K-edge region with the Arcus concept mission. In Section \ref{conclusions} we present the conclusions. 

\section{The laboratory experiment}
\label{lab}

\subsection{The dust samples}
\label{samples}

The dust samples used in this work are laboratory analogues of silicates and oxides of astronomical interest. The choice of samples reflects the astronomical silicate composition which has been discussed in previous studies (\citealt{Jaeger1998}, \citealt{Kemper2004}, \citealt{Chiar2006}, \citealt{Min2007}, \citealt{Olofsson2009}, \citealt{Draine1984}). In Table \ref{tab:samples} we present the list of the samples. The first column shows the reference number of each sample. In the second and third columns we present the compound name and its chemical formula and in the last column its form (i.e. crystalline or amorphous). We note that in this paper we use amorphous as a collective term in order to describe all non-crystalline materials. Our amorphous samples are glassy, the structure of which may still show a short-range order of atoms. However, Si-K edge spectra of our amorphous samples show a smooth profile, significantly distinct from the crystal ones (\citealt{Zeegers2019}). \\
 Most of the compounds have already been presented in previous works where the Mg, Fe, and Si K-edges have been measured in the laboratory using the Synchrotron facilities (\citealt{Zeegers2017,Zeegers2019}, \citealt{Rogantini2018}). Among the samples, four of them are natural (samples 2, 7, 9, 16 in Table \ref{tab:samples}) and six are commercial products (samples 11, 12, 13, 15, 17, 18). The rest of the silicate samples are synthesized in the laboratories at the Astrophysikalisches Institut and Universitats-Stenwarte (AIU) and Osaka University (for details see \citealt{Zeegers2019}).  \\
Our list of 18 samples contain pyroxenes, olivines and oxides. The silicate samples consist of different Mg/(Mg+Fe) ratio, between 0.5 and 0.9 and can be found in both amorphous and crystalline form. We also include magnesium pure compounds such as forsterite and enstatite (sample 7, 8, 11) and the iron pure compound named fayalite (sample 10). Additionally, we have 3 different types of quartz (samples 16, 17, 18), one in crystalline form and two in different stages of amorphisation.  \\

\begin{table*}[ht]
\begin{minipage}[t]{\hsize}
\setlength{\extrarowheight}{3pt}
\centering
\caption{List of dust samples.}
\newcommand{\head}[1]{\textnormal{\textbfh{#1}}}
\begin{tabular}{cccc}
  \hline
  \hline
Sample number & Compound name & Chemical Formula & Form\\
  \hline
  1 &  Olivine & $ \rm MgFeSiO_{4}$ &  amorphous \\
  2 &  Olivine & $ \rm Mg_{1.56}Fe_{0.4}Si_{0.91}O_{4}$ &  crystalline \\
  3 &  Pyroxene& $ \rm Mg_{0.9}Fe_{0.1}SiO_{3}$ & amorphous \\
  4 &  Pyroxene & $ \rm Mg_{0.9}Fe_{0.1}SiO_{3}$ &  crystalline \\
 5 &  Pyroxene & $ \rm Mg_{0.6}Fe_{0.4}SiO_{3}$ &  amorphous \\
 6 &  Pyroxene &  $ \rm Mg_{0.6}Fe_{0.4}SiO_{3}$ &  crystalline \\
 7 & Enstatite & $ \rm MgSiO_{3}$ &  crystalline \\
  8 & Enstatite & $ \rm MgSiO_{3}$&  amorphous \\
 9 & Hypersthene & $ \rm Mg_{1.502}Fe_{0.498}Si_{2}O_{6}$ &  crystalline \\
 10 &  Fayalite & $ \rm Fe_{2}SiO_{4}$ &  crystalline \\
 11 &  Forsterite & $ \rm Mg_{2}SiO_{4}$ & crystalline \\
 12  &  Spinel & $ \rm MgAl_{2}O_{4}$ &  crystalline \\
13  &  Corundum & $ \rm Al_{2}O_{3}$ &  crystalline \\
 14 & Pyroxene & $ \rm Mg_{0.75}Fe_{0.25}SiO_{3}$ &  amorphous \\
 15 & Magnetite & $ \rm Fe_{3}O_{4}$ &  crystalline \\
 16 &  Quartz & $ \rm SiO_{2}$ &  crystalline \\
 17 &  Quartz & $ \rm SiO_{2}$&  amorphous \\
 18 &  Quartz & $ \rm SiO_{2}$ &  amorphous\\
\hline
\hline
\end{tabular}
\end{minipage}
\label{tab:samples}
\end{table*}

\subsection{The Electron Energy Loss measurements}
\label{eels}

We perform laboratory measurements of dust in the oxygen K-edge using the STEM (Scanning Transmission Electron Microscope) facility at the University of Cadiz. Electron energy loss spectroscopy (EELS) is a technique which measures the change in the kinetic energy of electrons after the interaction with the sample. This provides structural and chemical information of the material studied. The interaction between the incident electron beam and the materials, in our case the dust samples, results in electron energy loss. A typical energy loss spectrum contains many features and is separated into three spectral regions. The so-called \textit{zero-loss} peak represents electrons that are transmitted without suffering measurable energy loss. The \textit{low-loss} spectrum corresponds to electrons that have interacted with the weakly bound electrons of the atoms and finally the \textit{core-loss} region which contains the electrons that have interacted with the tightly bound core electrons (\citealt{Egerton}). The \textit{core-loss} region contains the information on the fine structures of dust, also referred to as Energy Loss Near Edge Structures (ELNES). Both ELNES and XAFS present exactly the same spectral shape under the dipole approximation. \\
For the purpose of this experiment we use the Titan Cubed Themis 60-300 microscope in the STEM mode, operating at 200 keV. In parallel, high-spatial resolution EELS experiments were performed with the spectrum-imaging mode which allows the correlation of analytical and structural information on selected regions of the studied material. In this way we compare the information of the spectra with those of the spatial distribution of the sample. These supporting information experiments were acquired in Dual EELS mode using an energy dispersion of 0.25 eV, 50 pA probe current, and 50 ms acquisition time per EELS spectrum (\citealt{Manzorro}).\\
The STEM mode of the Electron Microscope enables us to scan pixel by pixel the dust samples. The resulting spectrum will be the average spectrum of each pixel. To improve the signal-to-noise ratio of our data we obtain 5-10 measurements of individual selected regions of interest on the dust sample. \\

\subsection{Analysing the laboratory data}
\label{eels_alanysis}
\begin{figure} [ht]
\begin{center}
\includegraphics[width=0.5\textwidth]{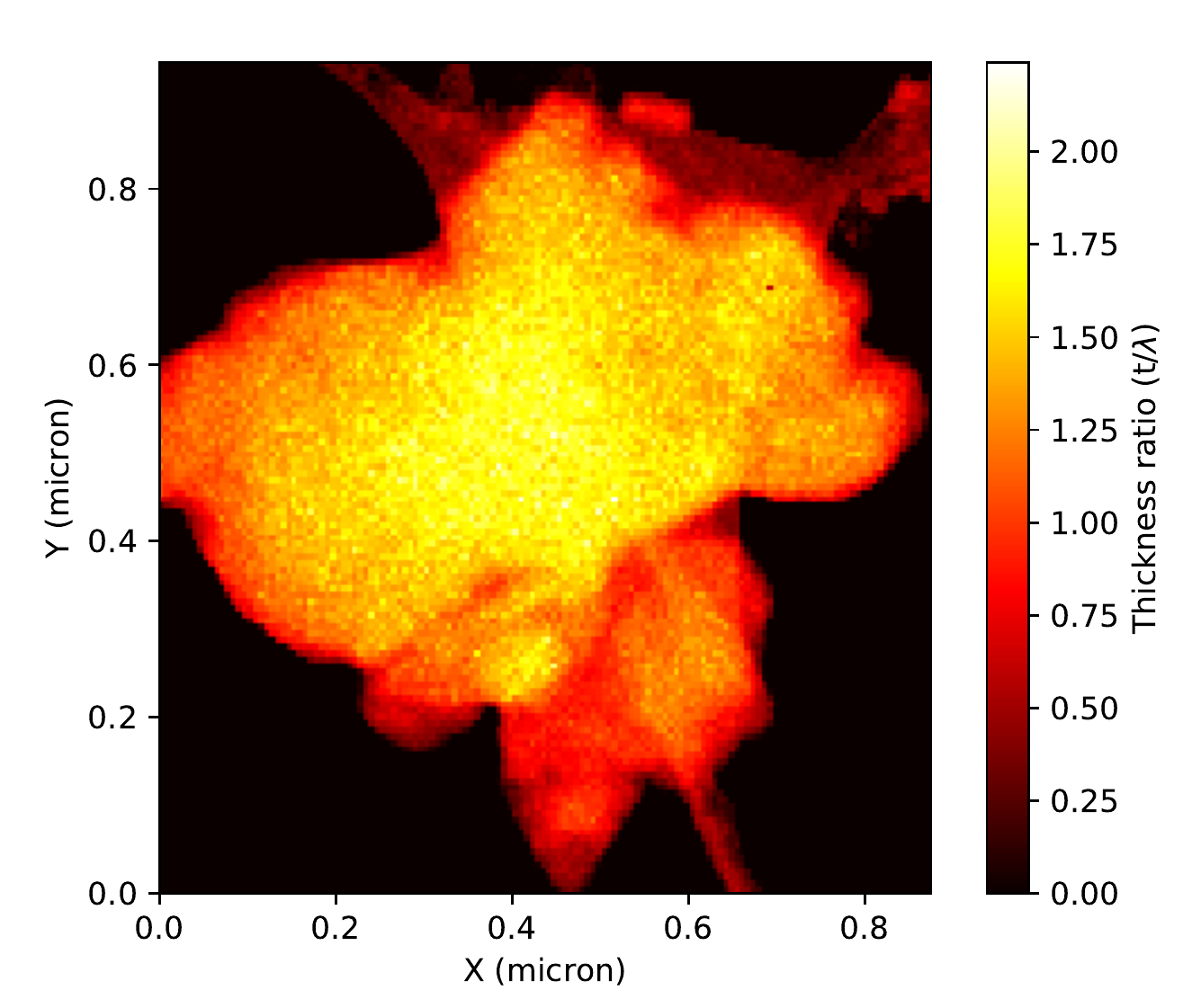}
\label{fig:grain}
\caption{Example of the region of interest on the dust sample obtained from laboratory data. The colourbar of the image indicates the thickness ratio $t/\tilde{\lambda}$, where $t$ is the thickness and $\tilde{\lambda}$ the mean free path of the inelastic scattering.}
\end{center}
\end{figure}

In this section we describe step by step the analysis of the raw laboratory measurements. We use the $\textit{python}$ library \textit{hyperspy}\footnote{https://hyperspy.org/} to convert the laboratory data to a python-readable format.  \\
- \textit{Thickness selection}: The dust samples have an intrinsic thickness variation. In the EELS, as the thickness of the area under study increases, the strong interaction of the primary electrons within the grain will result in multiple scattering events. This tends to reduce the signal-to-noise ratio of the EELS edges. To this end we need to apply a selection criterion. We select the part of the grain which belongs to thinner regions and then we consider the total spectrum of this part (see Fig. \ref{fig:grain}). In most cases, we exclude the part of the grain with $t/ \ \tilde{\lambda}$ >1, where $t$ is the thickness of the sample and $\tilde{\lambda}$ the mean free path of the inelastic scattering.\\
- \textit{Spectrum alignment}: An important correction in the EELS measurements is the alignment of the energy axis of the spectrum. For this reason we obtain the so-called $\textit{low-loss}$ spectrum which contains the zero loss peak (\citealt{Egerton}). The zero loss peak is a sharp peak with an energy loss of zero appearing in an EELS spectrum. We obtain this spectrum for every measurement. \\
- \textit{Background subtraction}: Next, we remove the EELS background by fitting a polynomial to the data.  \\
- \textit{Testing energy calibration uncertainties}: In laboratory experiments one needs to be careful about the absolute energy of the measured edges. It is possible that the calibration of the instrument is not perfect and therefore a verification of the absolute energy values has to be done. Therefore, we test the position of the oxygen edges with a reference feature that appears in our measurements. This narrow feature is a pre-peak of the oxygen edge and it is associated with an electron transition into molecular oxygen ($\rm O_{2}$), which is thought to evaporate due to the impact of the incident electron beam and the sample. We adopted a value of 530.5 eV for the absolute energy of the $\rm O_{2}$ transition (\citealt{Jiang}, \citealt{Garvie}). 

\subsubsection{Principal Component Analysis}
\label{pca}
We perform a Principal Component Analysis (PCA) to further test if the samples are homogeneous in composition and thickness. The original idea of PCA was established by Karl Pearson in 1901 \citep{pca}. The PCA is a statistical procedure that uses an orthogonal transformation to find inhomogeneity in a set of data. It decomposes a dataset into a set of orthogonal eigenvectors or principal components. When the PCA is applied to a set of spectra, the result is a set of spectral components which describe the variability of the data. The first principal component is the eigenvector that corresponds to the data with the maximum variation. The second eigenvector will show the second highest variation of the data and it is orthogonal to the first one. Every principal component will reveal the characteristic features of the spectrum in order of variation. \\
The dust grains may present inhomogeneities. Using the PCA, we cross check if the selected area on the grain is homogeneous. If it is not homogeneous we repeat the analysis for this specific measurement. This procedure is useful in order to obtain robust results and avoid artifacts in the spectra.  

\section{From laboratory measurements to cross sections} 
\label{labtocs}

From the EELS experiment we obtain the information about the fine structure of dust for each sample. Next, we need to calculate the extinction cross section in order to adopt the laboratory measurements for astronomical data analysis. The methodology used by us has been described extensively in previous works of \citet{Zeegers2017} and \citet{Rogantini2018}.
Here, we discuss the main steps of this analysis, adjusted for the oxygen K-edge.\\

\begin{figure} [ht]
\begin{centering}
\includegraphics[width=0.5\textwidth]{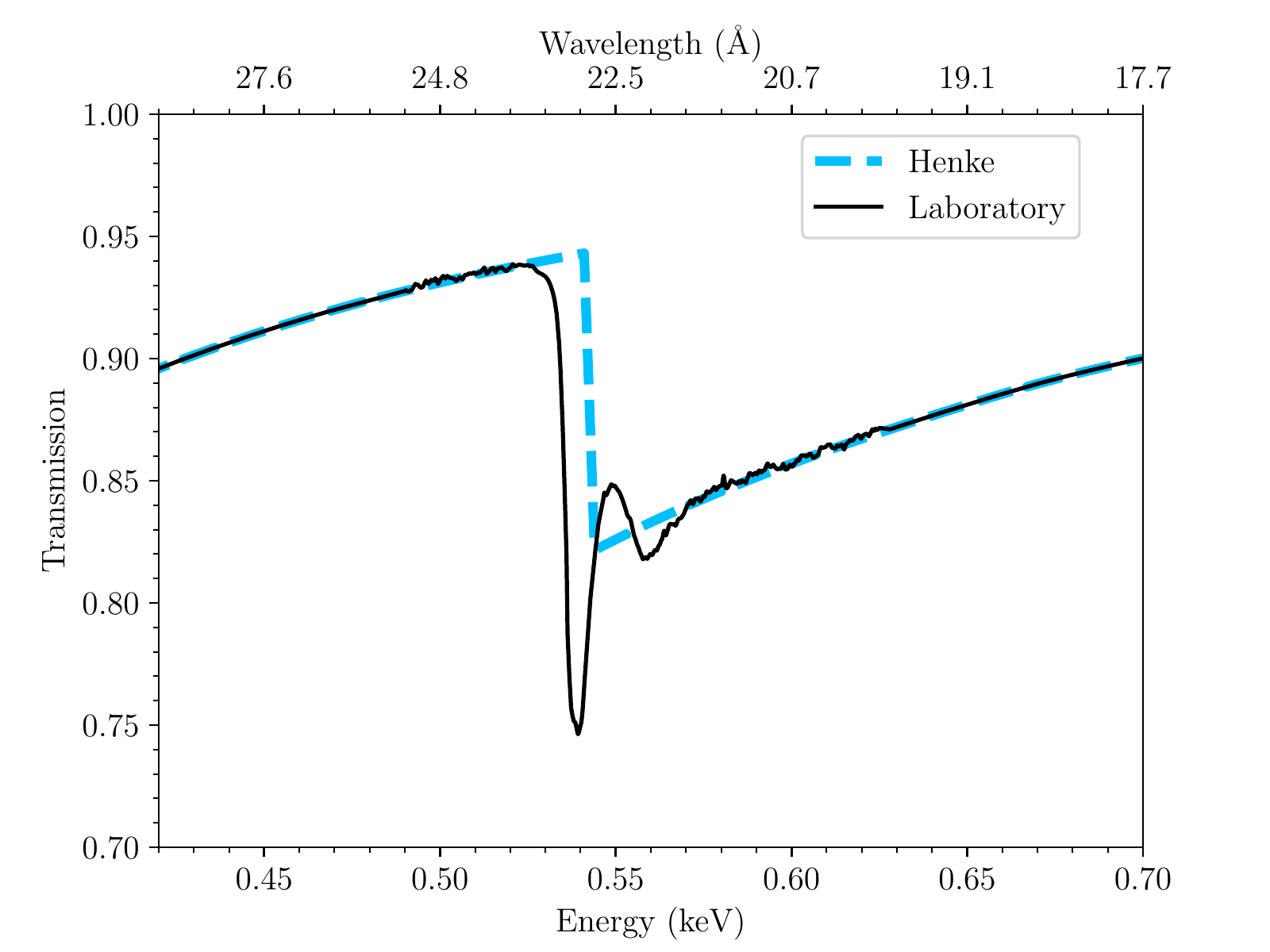}
\label{fig:henke}
\caption{Transmission spectrum in the oxygen K-edge using of amorphous olivine dust compound (black solid line). The blue dashed line indicates the tabulated data found in \citet{henke1993}, which were used to fit the pre-edge and post-edge of the data.} 
\end{centering}
\end{figure}
When the X-ray light pass through a material, it can be either transmitted, absorbed, or scattered. The transmittance of a material is the amount of light that is transmitted and it is described by the Beer-Lambert law:

\begin{align*}
T=\frac{I}{I_{o}}=e^{\mu\cdot x}
\end{align*}

\noindent where $I$ is the transmitted and $I_{o}$ the incident light intensity. In the above equation $\mu$ is the attenuation coefficient in $\mu \rm m^{-1}$ which depends on the properties of the material and $x$ describes the depth of the radiation in the material, i.e. the thickness of the sample (in $\mu{\rm m}$). The thickness should be 2-3 times below the attenuation length to avoid the thickness effects which can reduce the XAFS amplitude (\citealt{Parratt}, \citealt{Bunker}). For the attenuation length we use the tabulated values obtained from the Center for X-Ray Optics at Lawrence Berkeley National Laboratory (CXRO)\footnote{http://www.cxro.lbl.gov/}. In this work, we choose a thickness of 0.05 $\mu$m for all our samples, a necessary value in order to calculate the optical constants. This is consistent with the mean value of the EELS thickness of our samples which varies from $\sim$ 0.01 $\mu$m to $\sim$ 0.09 $\mu$m.\\
In Figure \ref{fig:henke} we present the transmittance of an olivine compound obtained using the EELS. The pre-edge (< 0.525 keV) and the post-edge (> 0.575 keV) have been fitted to the tabulated values from CXRO \citep{henke1993}.

\subsection{Optical constants}
\label{opticalconst}

In order to calculate the total extinction cross section of our samples we need to determine first the optical constants of the complex refractive index which is defined as:

\begin{align*}
m=n+ik
\end{align*}

The refractive index is a complex number that describes how the light propagates into a material. The real part shows the dispersive behavior on the incident light and the imaginary part its absorption. In the literature there are different notations to describe the optical constants \citep{Rogantini2018}. 

\begin{figure} [ht]
\begin{center}
\includegraphics[width=0.48\textwidth]{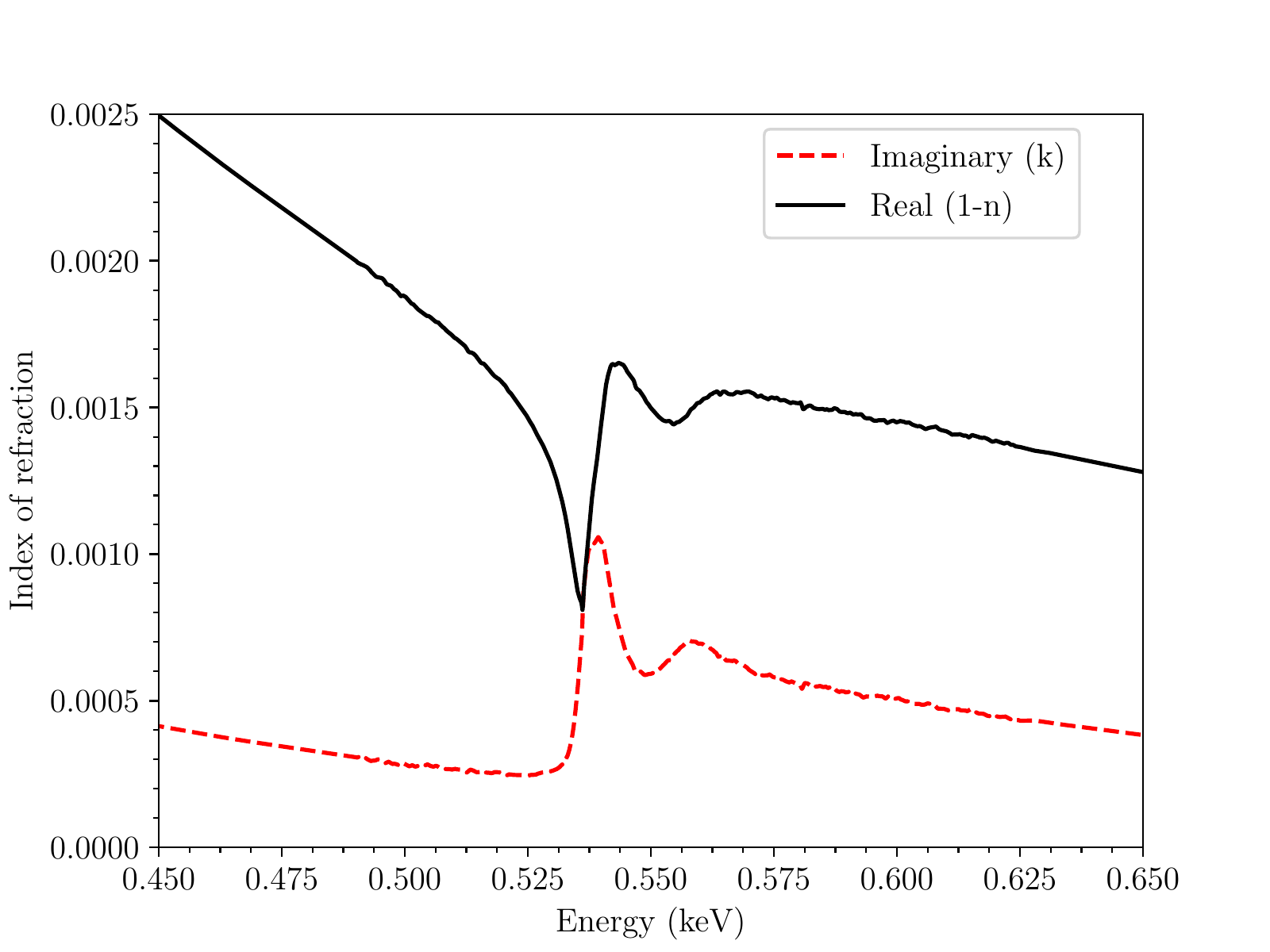}
\label{fig:nk}
\caption{Real and imaginary part of the refractive index calculated using the Kramers-Kronig relations for the amorphous olivine.}
\end{center}
\end{figure}

The imaginary part $(\it k)$ can be calculated directly from the data and the thickness of each sample from the equations:

\begin{align*}
k=\frac{\mu \lambda}{ 4 \pi} && \mu=-\frac{lnT}{x} 
\end{align*}

\noindent where $\lambda$, represents the wavelength of the incident X-ray and $T$ the transmittance. Knowing the imaginary part of the refractive index, we use the Kramers-Kronig relations to calculate the real part. The Kramers-Kronig relations are mathematical relations that connect the real and imaginary parts of any complex function. These relations are used to calculate the real part from the imaginary part (or vice versa) of functions in physical systems. Here, to calculate the real part we use the python package \textit{kkcalc}\footnote {https://pypi.org/project/kkcalc/}. This tool uses the optical constants in terms of atomic scattering factors, $f_{1}$ and $f_{2}$ which are described in \citealt{Watts14} (see also \citealt{Rogantini2018}). In Figure \ref{fig:nk} we present the optical constants for the amorphous olivine (compound \# 1). 

\subsection{Total extinction cross section}
\label{csext}

When the optical constants $n$ and $k$ have been calculated, one can proceed to the calculation of the total extinction cross section. 
To calculate the extinction cross section we apply the Mie theory \citep{Mie}. Therefore, we use the python module \textit{miepython} \footnote{https://github.com/scottprahl/miepython} which is the python version of MIEV0 code described in \citet{Wiscombe}. \\
The Mie code calculates the extinction and scattering efficiency factors ($Q$).
Then, the scattering, extinction and absorption cross sections are calculated from the efficiency factors of each grain of radius $r$ ($C= \rm \pi \it r^{2} Q$). However, the interstellar dust grains can be found in different sizes in the interstellar medium. To obtain the total scattering, extinction and absorption cross sections one needs to integrate over a grain size distribution:

\begin{align*}
\sigma (\lambda)= \int_{a_{-}}^{a_{+}}  Q \cdot n(r) dr
\end{align*}

\noindent where $n(r)$ is the size distribution. We use the Mathis-Rumpl-Nordsieck model (MRN, \citealt{Mathis}):

\begin{align*}
n(r)dr=A_{i}n_{H}r^{-3.5}dr
\end{align*}

\noindent where $n_{H}$ is the number density of H atoms and $A_{i}$ is a normalisation constant which depends on the type of dust. The normalisation constant is calculated using the equations described in \citet{Mauche}, and \citet{Hoffman}. 
For the purpose of this paper, we calculate the extinction cross section to be implemented into the \amol model of the software package SPEctral X-ray and UV modelling and analysis, \spex, version 3.05.00\footnote{\url{http://doi.org/10.5281/zenodo.2419563}} \citep{Kaastra1996, Kaastra2018}. \\

\begin{figure} [ht]
\begin{centering}
\includegraphics[width=0.5\textwidth]{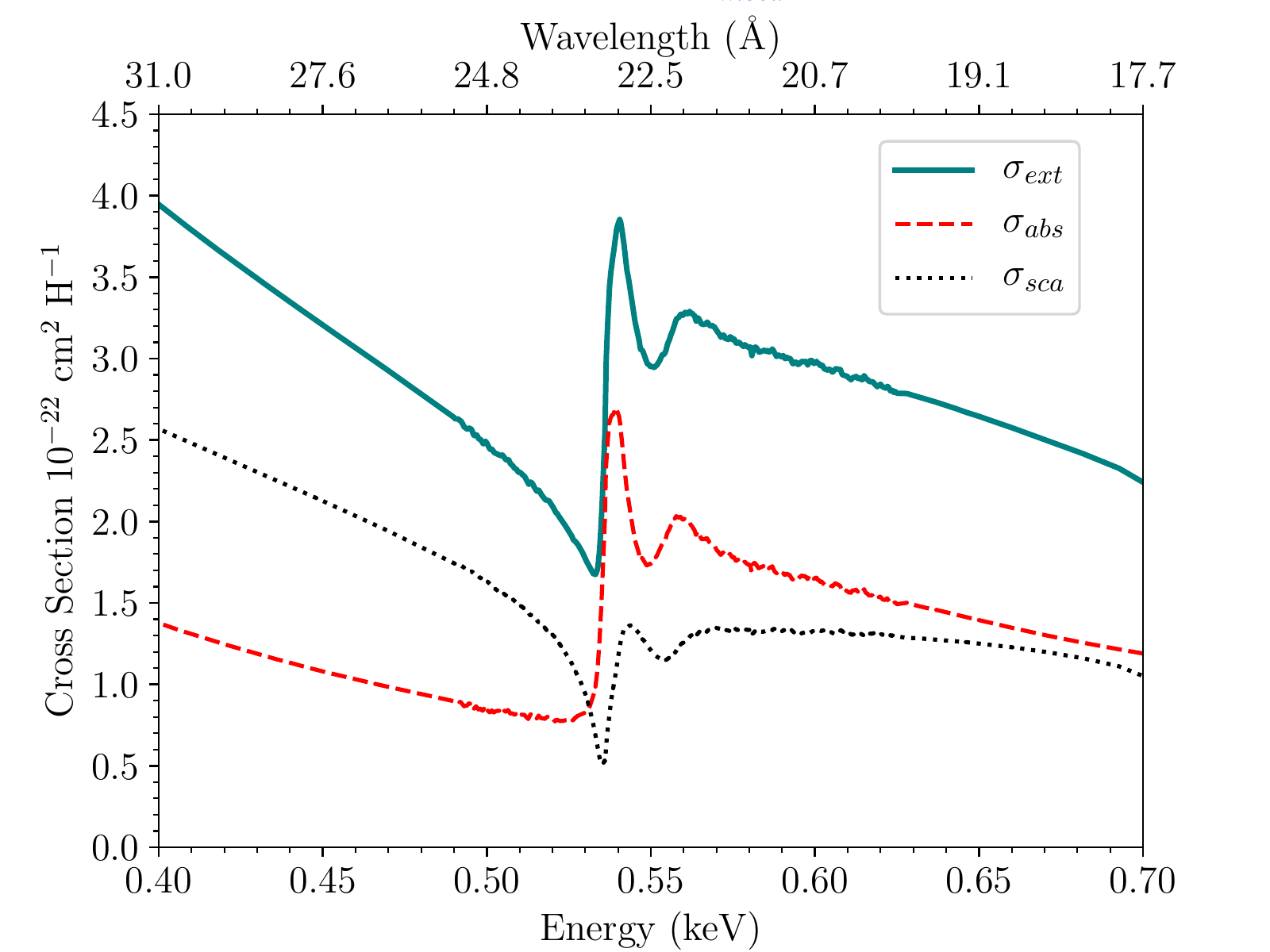}
\label{fig:cs3}
\caption{Calculated extinction ($\sigma_{\rm ext}$), absorption ($\sigma_{\rm abs}$) and scattering ($\sigma_{\rm sca}$) cross section of dust per hydrogen nucleus. The figure shows the amorphous olivine composition in the oxygen K-edge region.}
\end{centering}
\end{figure}

In Fig. \ref{fig:cs3} we present the calculated extinction, absorption and scattering cross sections for amorphous olivine. In Appendix \ref{appendix} we present the extinction cross sections for all compounds used in this work. The oxygen compounds measured in the laboratory show a smooth absorption profile. Contrary to the Si K- edge \citep{Zeegers2019} the differences among the different types of silicates are negligible. However, we note a difference between the amorphous and crystalline structure of the same compound. This effect is due to the different configuration of the atoms in the grain.

\section{Application to the astronomical data}
\label{astodata}

X-ray spectra of X-ray binaries can be used to study the interstellar medium in several lines of sight in the Galaxy through the absorption features of dust and gas.
Thus, we implement the laboratory data into a complete spectral model and apply them to Cygnus X-2, an ideal X-ray target to study the composition of the interstellar oxygen. For this analysis we use data from the Reflection Grating Spectrometer (RGS) on board the \xmm satellite \citep{denHerder}. The RGS has a resolving power of $R=\frac{\lambda}{\Delta \lambda } \gtrsim 400$ and an effective area of $\rm \sim 45 \ cm^{2}$ around the oxygen K-edge.

\subsection{Data reduction}
\label{datared}

 We reduce the \xmm data using the Science Analysis Software, $\textit{SAS}$ (ver.16). First, we run the $rgsproc$ command to create the event lists. Then, we filter the RGS event lists for flaring particle background using the default value of 0.2 counts/sec threshold. We also exclude the bad pixels using keepcool=no in the SAS task $rgsproc$. We use in total 4 long observations, listed in Table \ref{tab:obs}. In bright sources, some areas of the grating data may be affected by pile-up. In order to recognise the spectral regions not affected by pile-up one can compare the first and the second order for each grating (\citealt{Costantini2012}). We have negligible pile up in all the observations used in this paper. The spectral shape does not significantly vary through different epochs and this allows us to combine the data using the SAS command $\textit{rgscombine}$ in order to obtain a higher signal-to-noise ratio.

\begin{table}[htbp]
\begin{minipage}[t]{\hsize}
\setlength{\extrarowheight}{3pt}
\centering
\caption{\xmm observation log}
\newcommand{\head}[1]{\textnormal{\textbfh{#1}}}
\begin{tabular}{ccc}
  \hline
  \hline
  Obs. ID &  Obs. Date& Exp. Time (ks)\\
  \hline
  0303280101  &  14-06-2005 & 32   \\
  0111360101 &   03-06-2002 & 22  \\
  0561180501 &   13-05-2009 &  24  \\
  0602310101 &   12-05-2009 &  86 \\
\hline
\hline
\label{tab:obs}
\end{tabular}
\end{minipage}
\tablefoot{The exposure time refers to the net value, before the background filtering. The filtering causes the total exposure (sum of all the observations) to be reduced by $\rm \sim 40 ks$.}
\end{table}

\subsection{Spectral fitting}
\label{fitting}

\begin{figure*} [htbp]
\begin{centering}
\includegraphics[width=0.8\textwidth]{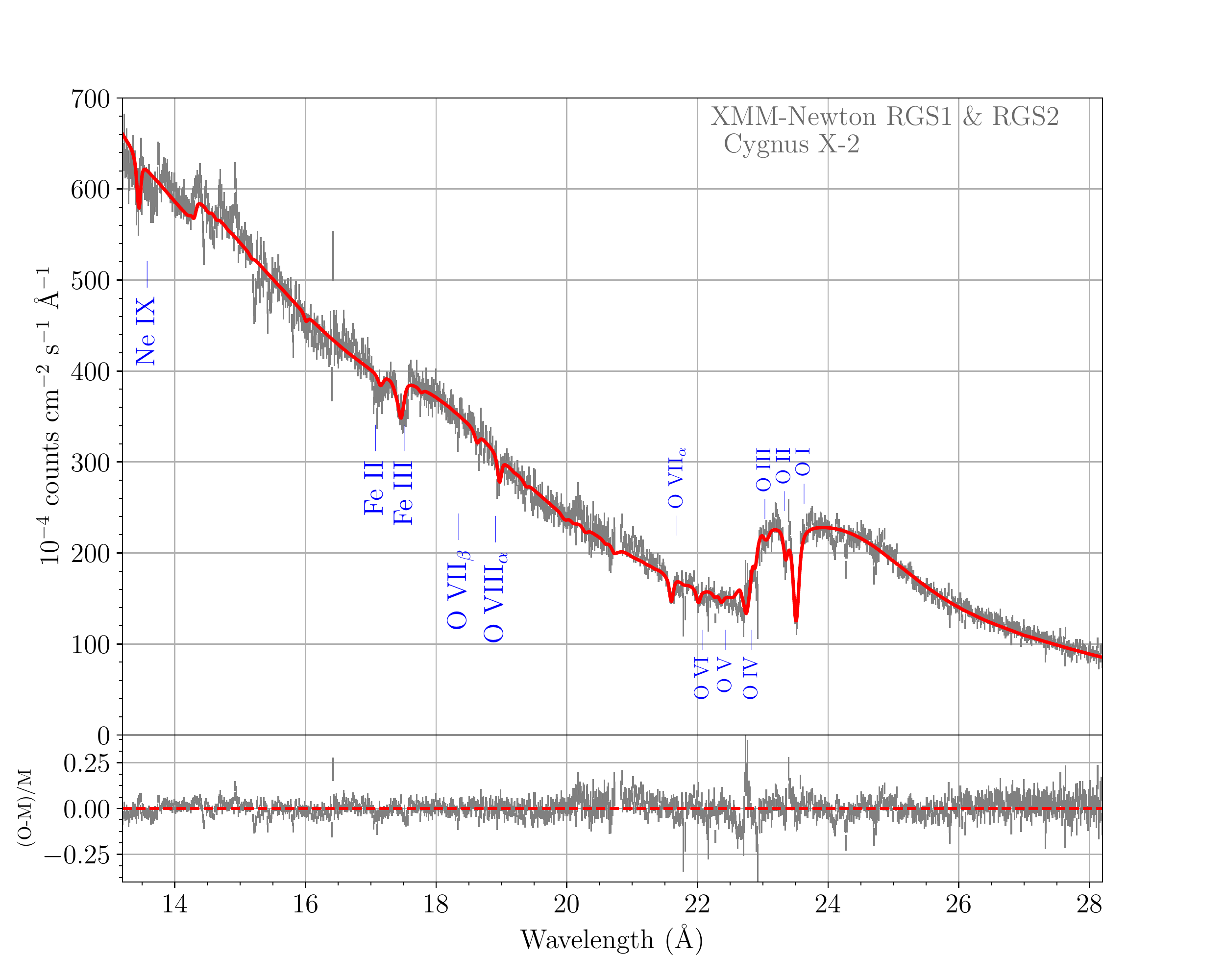}
\label{fig:continium}
\caption{Top panel: \xmm RGS spectrum of the low-mass X-ray binary Cygnus X-2 and the best fit model. Lower panel: Residuals of the best fit model, where (O-M)/M is (Observed-Model)/Model. }
\end{centering}
\end{figure*}

In Figure \ref{fig:continium} we present the stacked RGS spectrum of Cygnus X-2. The spectrum displays narrow absorption features near the oxygen ($\sim$ 23 $\AA$), neon ($\sim$ 13.5 $\AA$), and iron ($\sim$ 17.5 $\AA$) edges, which are produced by ionised gas in the ISM. In this work we mainly focus on the neutral and mildly ionised gas. We fit the spectrum in the range of 10-35 $\AA$ using SPEX. To evaluate the goodness of our fit we adopt $C$-Statistics which is denoted as $C_{\rm stat}$ in the rest of the manuscript. \spex uses $C_{\rm stat}$ as described in \citet{Cash1979} but with an addition which allows an estimate of the goodness of the fit (for details see \citealt{Kaastra2017}). Also, in our analysis we use the proto-Solar abundance units of \citet{Lodders2009}.\\
The continuum emission of low-mass X-ray binaries originates from their accretion disk and corona, and may be described as a blackbody and a Comptonised emission component, respectively (see \citealt{Done2007}). As RGS provides a relatively narrow energy band, the full shape of the spectral energy distribution cannot be constrained. Thus, we use a simple power law model (\pow in \spex), which reproduces well the continuum shape in the RGS band.  \\

\subsubsection{The gas phase modelling}
\label{contfit}

We further adopt a neutral gas component (\hot model in \spex) to absorb the continuum. The $\hot$ model calculates the transmission of a plasma in collisional ionisation equilibrium (\citealt{dePlaa}). For a given temperature and set of abundances, the model calculates the ionisation balance and then determines all the ionic column densities by scaling to the prescribed total hydrogen column density. At low temperatures ($\sim$ 0.5 eV) the $\hot$ model is effectively representing the neutral gas, since this temperature is sufficiently low to obtain neutral species. In this model both the hydrogen column density and the temperature (eV) are free parameters. This provides a physical fit of the ISM because we can distinguish between the different temperature phases along this line of sight. \\
We need in total three $\hot$ components to fit the weakly and mildly ionised features. These correspond to three different ISM temperatures along this line of sight: a quasi-neutral gas component with $kT \rm  \sim$ 0.7 eV, a warmer component with $kT \rm \sim$ 2.5 eV and a hotter one with $kT \rm \sim $ 12 eV (Fig \ref{fig:oxygen}). The parameters of the fit are shown in Table \ref{tab:continium}. We limit the temperature range of the cold gas to a maximum of 0.7 eV in order to have the cold gas phase dominated by neutral oxygen. \\
The stacked spectrum of Cygnus X-2 presents additional features: some absorption-like features around 24-25 $\AA$ and a 'bump' around 24-26 $\AA$. The strongest absorption-like features are at 24.7 $\AA$ and at 24.3 $\AA$ and their positions do not correspond to lines from known ions. In fact, they correspond to absorption features in the effective area of RGS. We concluded that these dips are most likely due to calibration issues associated to bad pixels and therefore can be ignored. \\
Moreover, the 'bump' in the continuum around 24-26 $\AA$ is too narrow to be associated with any continuum components. A similar bump is also observed after the Fe L-edge (around 18 $\AA$). These excesses have been associated to broad line emission features (\citealt{Madej2010, Madej2014}). For the improvement of our modelling we fit this excess with Gaussian profiles in order to prevent residuals in the oxygen edge region from a badly-fitted continuum and uncertainties in the determination of the absorption parameters. Thus, we leave the interpretation on the nature of these 'bumps' open.


\begin{table}
\caption{Best fit parameters for Cygnus X-2. }
\label{tab:continium}      
\centering
\renewcommand{\arraystretch}{1.1}
\begin{tabular}{c c c}        
\hline\hline           
Component & Parameter & Value \\    
\hline\hline
 \pow & $\rm norm$    &   $88.4 \pm 0.5$   \\
		 &  $\Gamma$            &   $1.25 \pm  0.01$        \\
 \hline
 \hot \# 1  & $N_{\rm H}  $            &  $1.9 \pm 0.1      $   \\
		&   $kT$      &     $  0.7 \ (fixed) $ \\
\hot \# 2  &  $N_{\rm H}$       &  $0.05 \pm 0.01 $      \\
		&   $kT$     &     $2.5   \pm 0.9 $             \\
\hot \# 3 &  $N_{\rm H}  $           & $ 0.05 \pm 0.01 $  \\
		&   $kT$    &  $  11.7  \pm 0.4 $   \\
\hline                                   
\hline  
\end{tabular}
\tablefoot{ The symbol $kT$ represents the temperature in eV and the $N_{\rm H} $  is the column density in $\rm 10^{21} \ cm^{-2}$. The symbol $\Gamma$ is the photon index and the normalisation of the power law is in units of $10^{44}$ \ $\rm ph/s/keV \ at \ 1keV$.}                            
\end{table}

\begin{table}
\caption{Best fit parameters of the \slab component. The symbol $N_{\rm ion}$ represents the ionic column density.}
\label{tab:slab}      
\centering
\renewcommand{\arraystretch}{1.5}
\begin{tabular}{c c}        
\hline           
\hline
 ion &$\log N_{\rm ion} \ (\rm cm^{-2}) $ \\
 \hline
 \ovi         &          $15.4   \pm 0.1   $      \\
\ovii          &          $15.9   \pm 0.1  $      \\
\oviii         &  	$15.8	   \pm 0.1 $   \\
\hline                                   
\hline                          
\end{tabular}
\end{table}

\subsubsection{The oxygen edge region}
\label{oxygenedge}

\begin{figure} [htbp]
\begin{centering}
\includegraphics[width=0.5\textwidth]{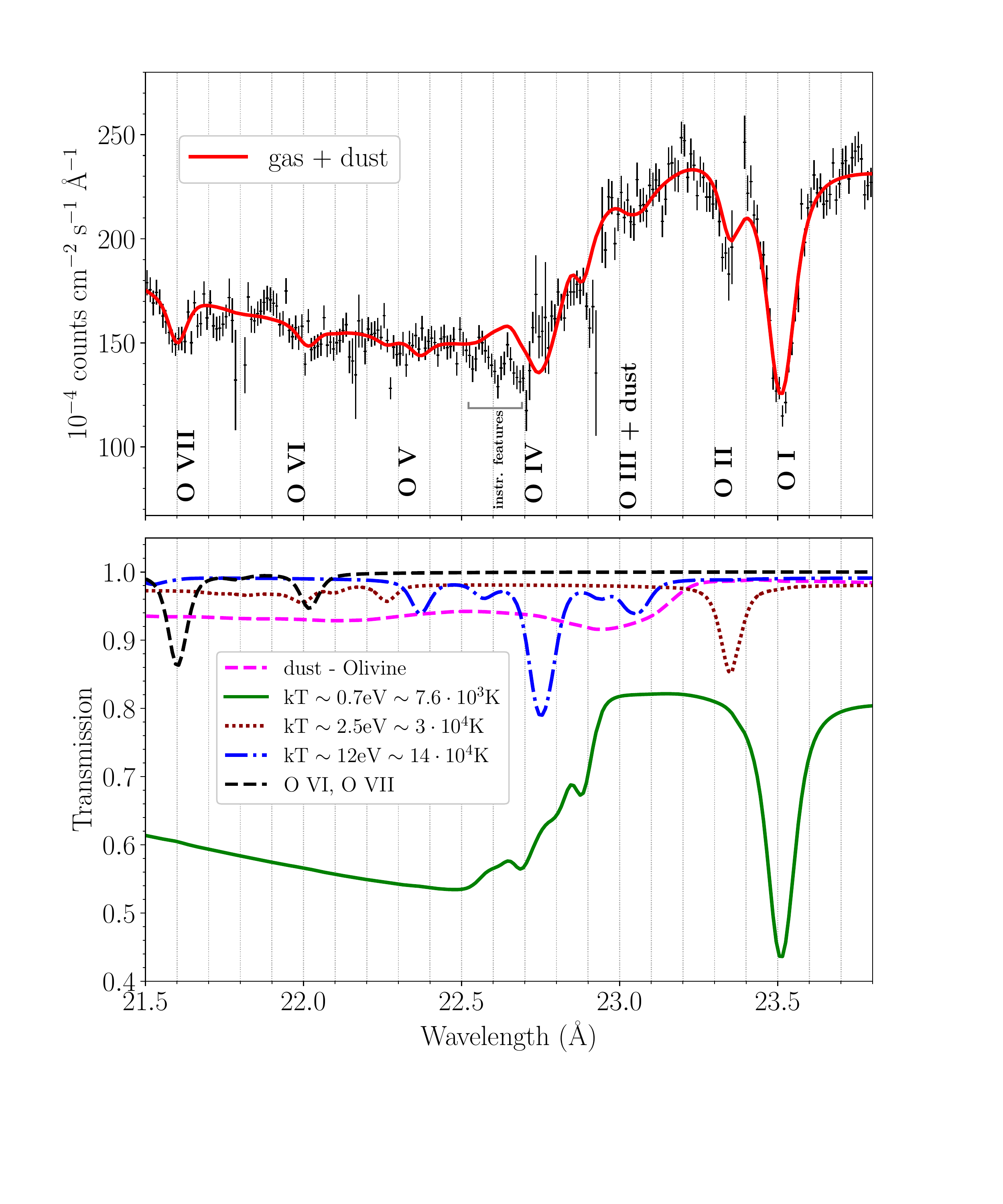}
\label{fig:oxygen}
\caption{Top panel: \xmm RGS spectrum of the low-mass X-ray binary Cygnus X-2. Best fit around the oxygen K-edge. Lower panel: The transmission of each component. The transmission of the colder component of the ISM has been multiplied by a factor of 2.5 for display purposes.}
\end{centering}
\end{figure}

 Here, we focus our analysis in the oxygen edge spectral region ($\rm 19-26 \ \AA$) which displays several sharp absorption features. We follow a similar fitting procedure for the oxygen edge as in \citet{Costantini2012}. The best fit is shown in the top panel of Figure \ref{fig:oxygen} and in the lower panel we present the contribution of each absorption component. The 0.7 eV gas (see Section \ref{contfit}) imprints the line from $\oi$ at 23.5 $\AA$ and is mimicing the neutral gas. At this stage, from our fit we obtain $C_{\rm stat}$= 1702 for 786 degrees of freedom ($d.o.f.$). At shorter wavelengths, the gas component with higher temperature of $kT$=2.5 eV produces the absorption by $\oii$ (23.36 $\AA$). A third component with slightly smaller contribution to the total fit takes into account the \oiii absorption line together with some weaker $\oiv$ and $\ov$ lines. At wavelengths 22 $\AA$, 21.6 $\AA$, 18.9 $\AA$ the lines of $\ovi$, $\ovii$ and $\oviii$ respectively, are the signature of an even hotter phase of the ISM. We take into account absorption by this hot gas using the \slab model in \spex. This model calculates absorption by a slab of optically-thin gas, where the column density of ions are fitted individually and are independent of each other (see Table \ref{tab:slab}). The additional \hot models and the \slab model improve our fit and we obtain $C_{\rm stat}$/$d.o.f.$ = 1380/772. \\ 
Finally, we test the effect of our new calculated dust extinction cross sections implemented into the \amol model of \spex. We let the column density of the \amol model free to vary as well as the depletion of the \hot model (hot \# 1, cold phase) of oxygen and iron. The depletion of silicon is fixed to 90\% according to the literature (\citealt{Rogantini2019}, \citealt{Zeegers2019}) and the depletion of magnesium is constrained to be at least 80\%.\\
In Figure \ref{fig:oxygen} we present the best fit using an olivine dust compound (amorphous olivine, compound \# 1 in Table \ref{tab:samples}). In the lower panel of Figure \ref{fig:oxygen} we present the transmission of the dust component (in magenta). As shown, the transmission of the dust covers a broad area of the oxygen K-edge spectral region from about 22.3 $\AA$ to 22.5 $\AA$. By adding the dust model we obtain $C_{\rm stat}$/$d.o.f.$= 1342/771.  \\


\subsection{The dust modelling}
\label{dust}

Our fit takes into account the contribution of dust using the newly calculated extinction models. The analysis above has been carried out using an amorphous (glassy) olivine composition ($ \rm MgFeSiO_{4}$). However, the ISM consists of a mixture of chemical elements. We have further tested different combinations of dust compounds from the list in Table \ref{tab:samples}, containing silicates and oxides, to search for the best fit combination which describes the interstellar dust composition in the oxygen edge. The \amol model in \spex allows us to test 4 different compounds at a time. The number of different combinations is given by the equation:

\begin{align*}
\rm C_{e,c}=\frac{e!}{c!(e-c)!}
\end{align*}

\noindent where $e$ is the number of the available edge profiles and $c$ the combination class (see also \citealt{Costantini2012}). In our fit we do not include the aluminium compounds (Table \ref{tab:samples}, compounds \# 12 and 13) since this element is outside the RGS energy band and it is believed that they exist in small quantities in the ISM. We fit with the remaining 16 compounds which gives 1820 different combinations. \\
To compare the different models we use a robust statistical approach. The standard model comparison tests (such as the $\chi^{2}$ Goodness-of-Fit test) cannot be used because our models are not nested (\citealt{Protassov}). Therefore, we perform the $\textit{Akaike Information Criterion}$ (AIC) which represents a robust and fast way to select the models which have more support from the data (\citealt{Akaike}), as explained in detail in \citet{Rogantini2019}. AIC gives the relative quality of statistical models for a given set of data. \\
We calculate the AIC difference ($\Delta AIC$) over all candidate models with respect to the model which has the lowest AIC value. The most significant model is the one which minimises the AIC value.
 Using the criteria of \citet{Burham}, the models with $\Delta AIC > 10$ can be ruled out. 
 From the total of 1820 models we obtain using the different combinations, about 590 of them were found with $\Delta AIC > 10$ and are therefore excluded. These models contain mostly Quartz (compounds \# 16, 17, 18 in Table \ref{tab:samples}). The rest of the models, with $\Delta AIC < 10$, are plausible to represent the data. In particular, models with $\Delta AIC < 2$ can fit equally well the oxygen edge. We obtain 311 different models with $\Delta AIC < 2$. We conclude that we do not have sufficient resolution to distinguish between the different dust samples. The modelling is further complicated by the relative similarity of the oxygen absorption dust profiles measured in the laboratory (Appendix \ref{appendix}).


\section{Discussion}
\label{discussion}

\subsection{The multi-phase gas in the ISM}
\label{multiphase}

The study of the ISM through the X-ray absorption features of bright background sources, such as X-ray binaries, is a unique tool to probe the different phases of the ISM. By studying the oxygen K-edge spectral region, we can constrain different gas temperatures by probing low and high ionisation gas, covering from \oi to \ovii. With our modelling, we find a column density of oxygen in the gas phase of $\sim 1.27 \times 10^{18} \rm cm^{-2}$ which corresponds to $(93\pm9)\%$ of the total column density. The remaining $(7\pm1)\%$ corresponds to solid phase (dust) with column density of $\sim 9 \times 10^{16} \rm cm^{-2}$. Considering the errors, if the percentage is lower or higher than 100$\%$ this implies that there is under or over abundance of oxygen, relative to the assumed abundances.\\
In our modelling the gas consists of 3 components (see Table \ref{tab:continium}). We have a cold gas component with temperature $ kT \rm =0.7 \ eV$ (or $T \rm \sim 8 \times 10^{3}\ K $), a warm component with $ kT \rm = 2.5 \ eV$ (or $ T \rm \sim 3 \times 10^{4} \ K$) and finally a hot component with $ kT \rm =12 \ eV$ (or $T \rm \sim 14 \times 10^{4}\ K$). 

\begin{table*}
\caption{Contributions to the total oxygen column density.}
\label{tab:contributions}      
\centering
\begin{tabular}{c c c c cc}        
\hline           
\hline
Phase & Constituent &  $N \rm _{o} (10^{18} cm^{-2}) $& $ \% N_{\rm o}^{\rm gas}$  &  $\% N_{\rm o}^{\rm total} $\\
 \hline
		 &      $\oi$                     &      $1.2  \pm 0.1 $      &     $95 \pm 9 $       &                  \\
Gas  &      $\oii$ - $\oiv$           &   $0.05 \pm 0.01 $      &    $4.0 \pm 0.8 $          &   $93   \pm 9 $            \\
		&      $\ovi$ - $\oviii$       & $0.020 \pm  0.001$          &      $1.00 \pm 0.01 $      &                  \\
 \hline
 Dust &     Silicates             & $0.09 \pm 0.01 $          &              &     $7   \pm 1    $     \\
\hline                                   
\hline                          
\end{tabular}
\tablefoot{ $N_{\rm o}$ represents the total column density, $ \%N_{\rm o}^{\rm gas}$ is the contribution of each gas constituent to the respective phase and $ \%N_{\rm o}^{\rm total}$ is the contribution of the different phases to the total column density. }
\end{table*}
In Table \ref{tab:contributions} we present the column density of each oxygen ion and its contribution to the total column density. The cold component produces mainly \oi, some of which is locked into dust. The warm component produces the low ionisation absorption line (\oii) and the hotter component produces the  \oiii, \oiv and \ov. The higher ionisation lines are reproduced using a \slab model in \spex. \\
Molecules such as carbon monoxide are expected to be found in the ISM, especially in molecular clouds. It has been proposed that the missing oxygen from the gas phase of the ISM could be locked up in molecules together with elements with a high cosmic abundance such as CO \citep{Jenkins2009}. To test the presence of carbon monoxide we use the cross section (\citealt{Barrus1979}) implemented into the \amol model in \spex. According to this model, such a feature is expected to be present around 23.2 \AA. We obtain an upper limit of $N_{\rm CO}< 3.2 \times 10^{16} \rm cm^{-2}$ which corresponds to less than $2\%$ of the total oxygen.

\subsection{Abundances and depletions}
\label{abundances}

From the best fit of Cygnus X-2 we find a column density of neutral hydrogen of $N_{\rm H} = (1.9 \pm 0.1) \times 10^{21} \rm cm^{-2}$. This is in agreement with the hydrogen column density along this line of sight \citep{HI4PI}. We find that $(7\pm1)\%$ of the oxygen is depleted into dust. \citet{Pinto2010} studied the interstellar medium along the line of sight of GS 1826-238 and found that $(10\pm2)\% $ of oxygen is contained within solids, such as silicates and water ice. Later, \citet{Costantini2012} studied the absorbed spectrum of 4U 1820-30 and found $(20\pm2)\%$ depletion. Our best fit indicates that the amount of oxygen in gas and dust is consistent with Solar composition. The results on the abundances and depletions are presented in Table \ref{tab:depletions}.\\

\begin{table}
\caption{Oxygen column densities and abundances.}
\label{tab:depletions}      
\centering
\renewcommand{\arraystretch}{1.1}
\begin{tabular}{cc}        
\hline           
\hline
$ N _{ \rm gas} $ & $(1.27 \pm 0.09) \times 10^{18} \rm cm^{-2}$ \\
$ N _{ \rm dust} $ &  $(0.09 \pm 0.01) \times 10^{18} \rm cm^{-2} $\\
$ A_{ \rm gas}$  & $ (6.1 \pm 0.6) \times 10^{-4} $\\
$A  _{ \rm dust}$ & $ (0.45 \pm 0.07) \times 10^{-4}$\\
$ A/A_\odot\ $   & $1.1 \pm 0.1$ \\
\hline 
\hline
\end{tabular}
\tablefoot{$N_{ \rm gas}$ and  $N_{ \rm dust}$ indicate the total column density of gas and dust respectively.  $ A_{\rm gas}$ and $A_{\rm dust}$ are the abundances of oxygen in gas and dust respectively (with respect to hydrogen). $ A/A_\odot$ is the total abundance (gas+dust) ratio in proto-Solar abundance units.}
\end{table}

\subsection{Comparison between the atomic databases for oxygen}
\label{atomic}

\begin{figure} [htbp]
\begin{centering}
\includegraphics[width=0.48\textwidth]{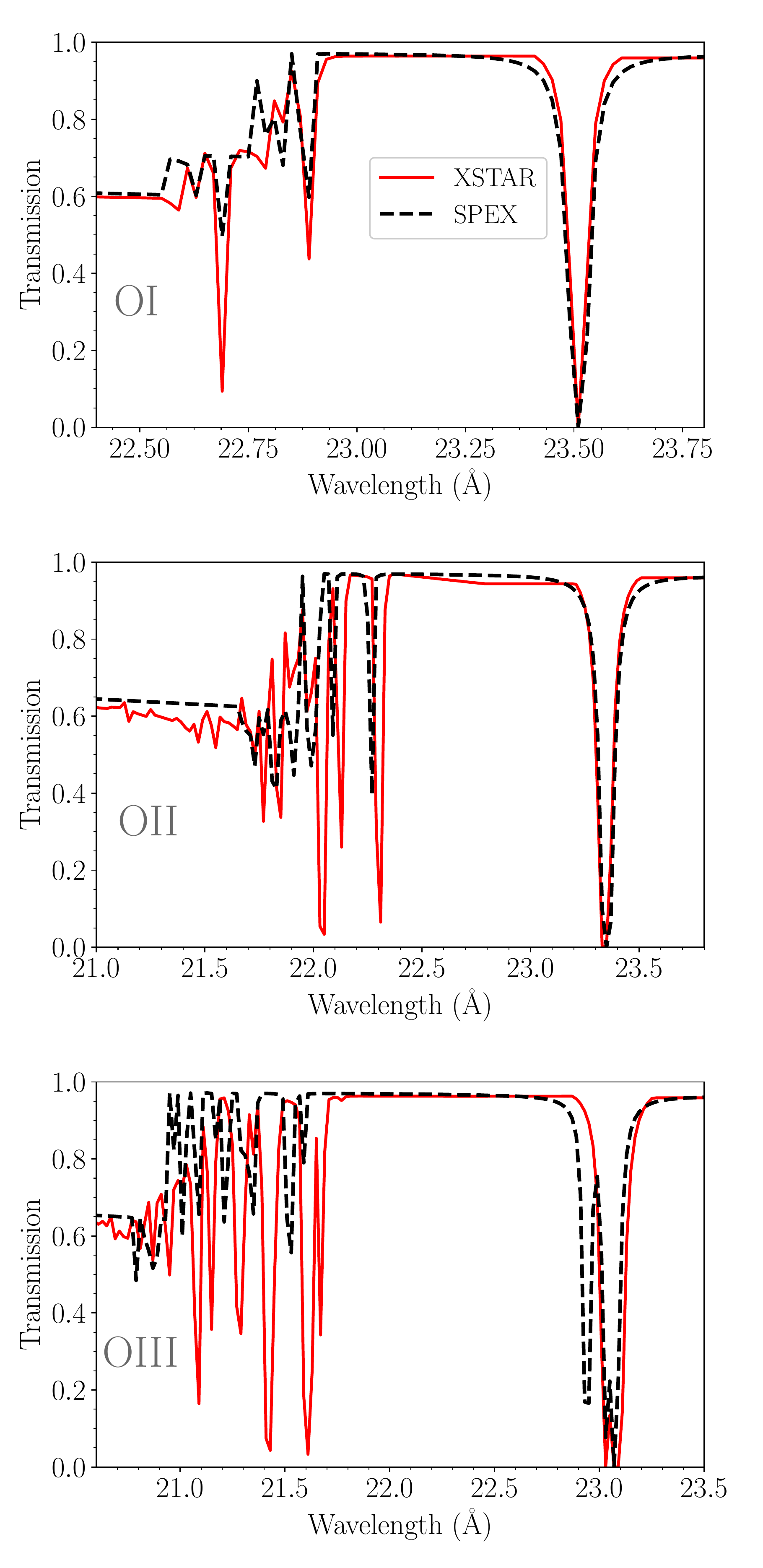}
\label{fig:atomic}
\caption{Comparison of the atomic databases of \spex and \xstar at the oxygen K-edge for $\oi$ (top panel), $\oii$ (middle panel), and $\oiii$ (bottom panel).}
\end{centering}
\end{figure}

\begin{figure} [htbp]
\begin{centering}
\includegraphics[width=0.5\textwidth]{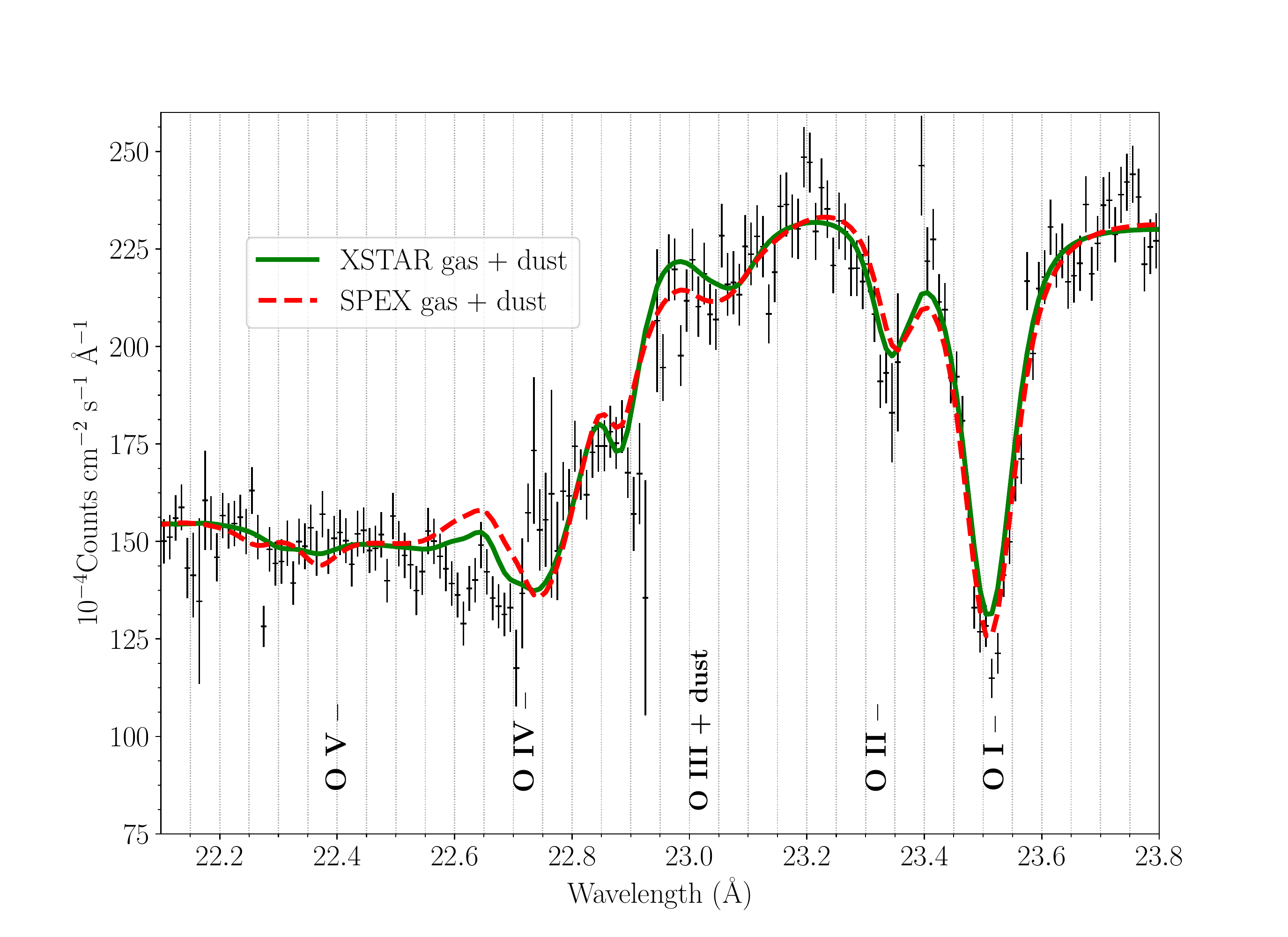}
\label{fig:atomicOx}
\caption{Fitting the oxygen K-edge region using different atomic databases for the oxygen (\oi, \oii, \oiii). The plot compares the fits using the \xstar and \spex atomic databases. The dust compound used here is the amorphous olivine.}
\end{centering}
\end{figure}

Accurate high-resolution X-ray spectroscopy and interpretation require state-of-art atomic databases. The oxygen K-edge region is one of the most important and complicated astrophysical spectral regions.
The available atomic X-ray databases for atomic oxygen are based on different cross section calculations. \\
Here, we compare the photoabsorption cross sections implemented into \spex (\citealt{Kaastra1996}) and \xstar (\citealt{Kallman2001}) for the atomic oxygen $\oi$, $\oii$ and $\oiii$. We further examine the effect of the different atomic databases in our fitting of the oxygen edge region of the interstellar medium. \\
Regarding \xstar, the $\oi$ photoabsorption cross section can be found in \citet{Gorczyca} and for $\oii$, $\oiii$ it uses the R-matrix calculations from \citet{Garcia2005}. \\
For the inner K-shell transitions of oxygen ions \spex uses the calculations provided by E. Behar (HULLAC calculations). For $\oi$ and $\oii$ the lines are shifted in order to match the \citet{Juett2004} observational values (see \citealt{Kaastra_OI}). Moreover, the oscillator strengths and transition rates for all $\oi$ lines have been updated with values computed by Ton Raassen using Cowan's code (\citealt{Cowan}, T. Raassen private communication). In Appendix \ref{spexlin} we present the wavelengths and oscillator strengths of the oxygen lines implemented in \spex. \\
In Figure \ref{fig:atomic} we present the comparison between the different atomic databases in the oxygen K-edge region. The top panel shows the model calculations for the atomic transitions of $\oi$. The cross sections of $\oi$ agree with each other regarding the wavelengths of the lines. The middle panel of Figure \ref{fig:atomic} refers to $\oii$ and the bottom panel to $\oiii$. For \oii and \oiii the absolute energy of the lines between \spex and \xstar is in disagreement. \\ 
We further test how the discrepancy between the atomic databases can affect the spectral fitting of Cygnus X-2. We substitute the atomic lines of \spex for $\oi$, $\oii$ and $\oiii$ with those of \xstar. This enables us to fit the lines of these three ions with the \xstar atomic data. In order to do that we use the \user model which allows external models into \spex. We fit the stacked spectrum of Cygnus X-2 using the lines for $\oi$, $\oii$, $\oiii$ from the \user model. We keep the same modelling as described in Section \ref{contfit}, as well as the \slab and the \amol model. The free parameter of the \user model in our case is the column density and is fitted independently for each individual $\oi$, $\oii$ and $\oiii$ ion. We keep the ratio of the column density of these three ions consistent with the ratio of the column densities and temperatures in the $\hot$ model in \spex in order to set a physical fit. In Figure \ref{fig:atomicOx} we present the best fit results. The \xstar database gives a better fit around 22.7 $\AA$. We have noticed that there is a weaker transition of $\oi$ in that region and using the \xstar database we obtain reduced residuals. In the rest of the spectrum we achieve a similar fit. 
 
\section{Prospects of the oxygen K-edge study with the Arcus-concept mission}
\label{arcusss}

\begin{figure*} [htbp]
\begin{centering}
\includegraphics[width=1\textwidth]{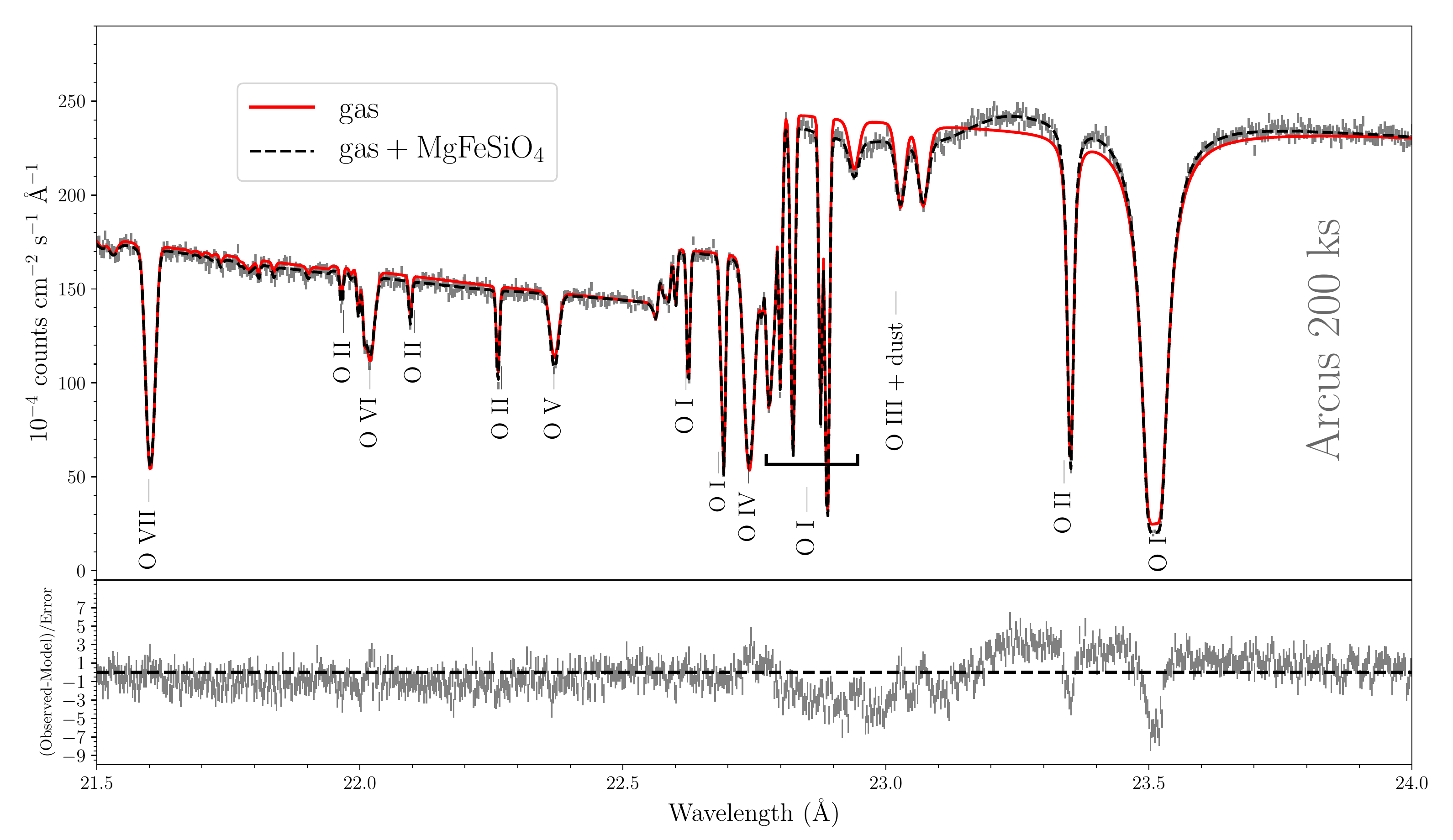}
\label{arcus}
\caption{Simulated data of Cygnus X-2 in the oxygen K-edge using the Arcus response matrix. Top panel: Simulated spectrum according to our best fit model (\ref{oxygenedge}) which includes dust (black dashed line). The red solid line shows the Arcus simulation without including a dust model. Bottom panel: Residuals of the fit with the gas-only model. }
\end{centering}
\end{figure*}

Future missions may provide us with suitable instruments to study the ISM in the soft X-rays (<1 keV). For example the Arcus-concept mission (\citealt{Arcus}) will provide higher resolution data with a resolving power of R $\sim$ 3000. \\
We perform an Arcus simulation in order to understand if we can put better constrains on the properties of the ISM and distinguish between the gas and solid phase. We use the \xmm RGS best fit model in the oxygen edge region described in Section \ref{oxygenedge} as a template to simulate an Arcus spectrum of Cygnus X-2. This model contains both gas and dust. In Figure \ref{arcus}, we present the Arcus simulation with an exposure time of 200 ks.  \\
We test the effects of including or excluding dust in our spectrum by fitting an only-gas model (red line). It is clear from the shape of the model around 23.2 $\AA$ that with Arcus we will better distinguish the effect of dust in the oxygen K-edge. Also, the wealth of individually resolved lines will place stronger constraints on the exact structure of the multiphase gas. In general, new X-ray missions such as ATHENA and XRISM and concept missions such as ARCUS will open up new frontiers to understand better the physics of the ISM (e.g.  \citealt{Rogantini2018}, \citealt{Costantini2019}, \citealt{Corrales2019}).  \\

\section{Conclusions}
\label{conclusions}

In this work we present a set of laboratory measurements in the O K edge region, and we aim to characterise the absorption properties of silicates and oxides, likely forming the interstellar dust content. We have measured 18 different dust species including silicates and oxides and we have calculated the dust extinction cross sections. We adopt the laboratory data for astronomical data analysis and in particular we implemented the dust models into the \spex X-ray fitting package. We further model the \xmm RGS spectrum of Cygnus X-2 using the existing models of gas in \spex and the new laboratory data. We focus our analysis in the oxygen K-edge region in order to study the dust and the properties of the multi-phase ISM along the line of sight to the source. We further discuss the effect of using different atomic databases for the atomic oxygen and finally we comment about new frontiers in the oxygen K-edge from the future concept mission of Arcus. Our main results are the following:\\

\begin{itemize}
	 \item From the oxygen K-edge we are able to study the multi-phase gas. We probe different gas temperatures and disentangle lower and higher ionised gas components, such as \oi and \oviii. 
	 \item The absorption spectrum of Cygnus X-2 shows the presence of gas and dust in the oxygen K-edge region. We find that $(93\pm9) \%$ of the total amount of oxygen is in the gas phase while a smaller amount of $(7\pm1) \% $ is found in dust. 
	 \item The oxygen abundance along the line of sight of Cygnus X-2 is consistent with proto-Solar values. This could be due to the fact that Cygnus X-2 is located further from the Galactic center compared to other X-ray binaries. 
	 \item To fully study the dust in the oxygen K-edge we need instruments with higher resolution and effective area such as the Arcus-concept mission. This will help to minimize the statistical noise and to better distinguish between the gas and solid phase.
    \item Accurate atomic databases are necessary in order to accurately study the oxygen region. The importance of the atomic data is highlighted in view of future X-ray missions. 
\end{itemize}

\begin{acknowledgements}
 We would like to thank our referee J. Nuth for the useful suggestions. IP, MM, DR and EC are supported by the Netherlands Organisation for Scientific Research (NWO) through The Innovational Research Incentives
Scheme Vidi grant 639.042.525. The Space Research Organization of the
Netherlands is supported financially by the NWO. The authors would like to thank E. Gatuzz for providing the cross section calculations of the atomic oxygen implemented into \xstar, J. Garcia for valuable suggestions on atomic data and J. Wilms, R. Smith for useful information about Arcus. We thank J. de Plaa for helping with the \user model in \spex and J. Kaastra for general instructions regarding \spex. We also thank M. Diaz Trigo for useful information regarding Cygnus X-2.  
\end{acknowledgements}

\vspace{-0.4cm}
\bibliographystyle{aa}
\bibliography{references}

\clearpage
\newpage

\appendix
\section{\\Dust extinction cross sections in the oxygen K-edge}
\label{appendix}
We present the calculated dust extinction cross sections in the oxygen K-edge. The cross sections were calculated from laboratory data of 18 dust samples with different chemical compositions. The resolution of the laboratory measurements is 0.25 eV.

\begin{figure}[ht]
\begin{centering}
\includegraphics[width=0.53\textwidth]{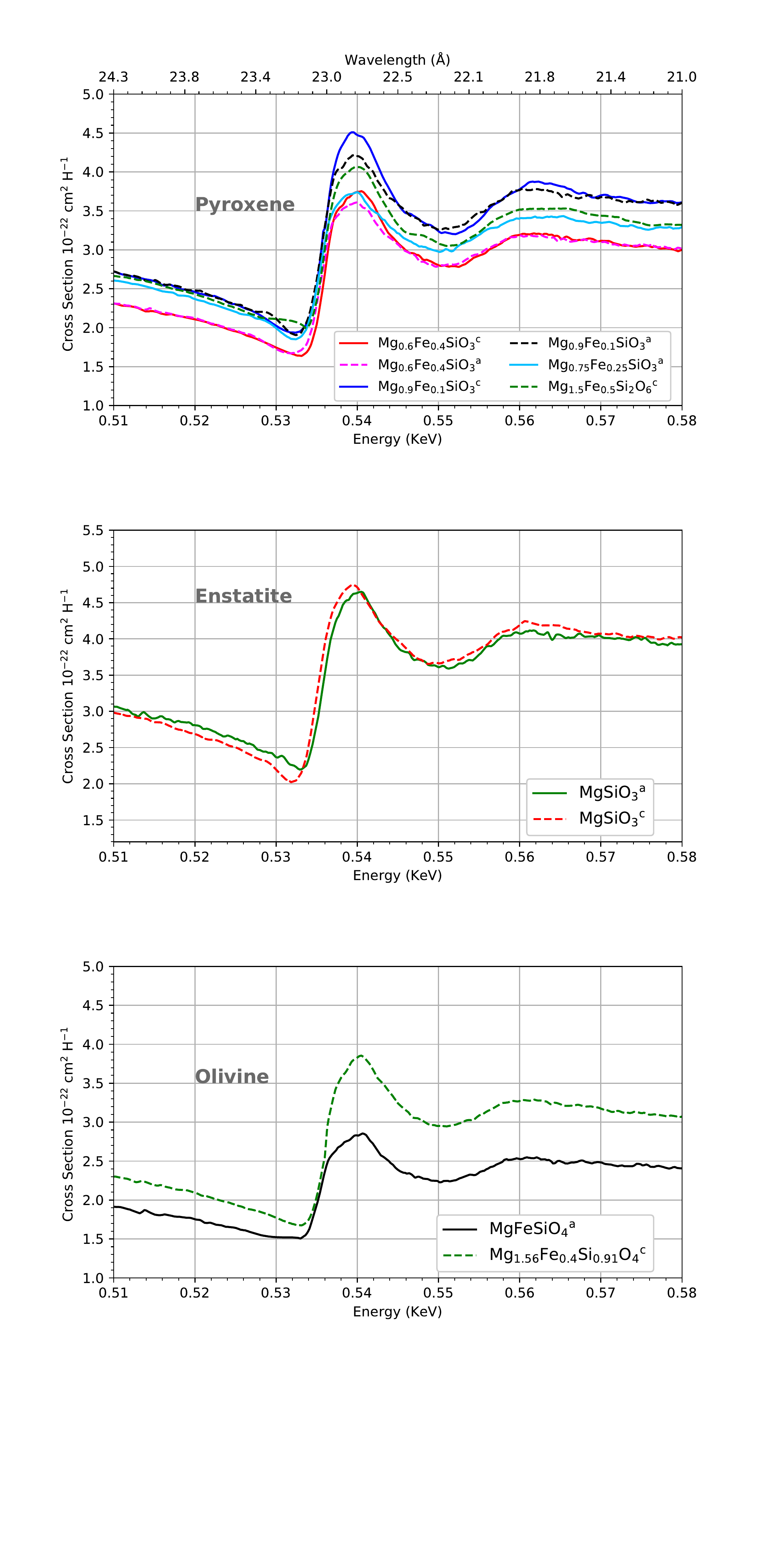}
\label{cs_all}
\caption{Calculated dust extinction cross sections. The symbol $a$ refers to amorphous compounds and $c$ to crystalline. }
\end{centering}
\end{figure}

\begin{figure}[ht]
\begin{centering}
\includegraphics[width=0.55\textwidth]{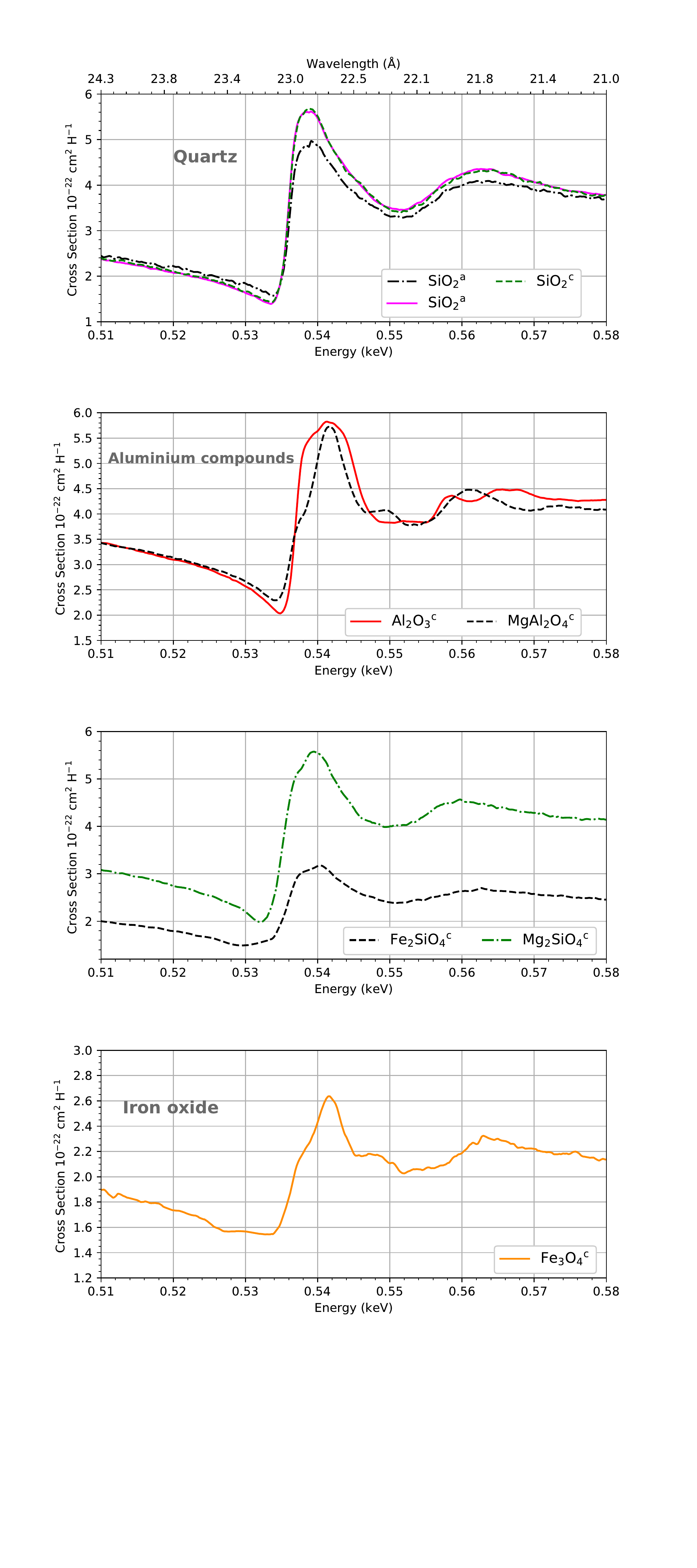}
\label{cs_all}
\caption{Calculated dust extinction cross sections. The symbol $a$ refers to amorphous compounds and $c$ to crystalline. }
\end{centering}
\end{figure}

\clearpage
\newpage

\section{Atomic oxygen lines in \spex}
\label{spexlin}

In the following tables we present the wavelength, energy and oscillator strength of the atomic oxygen lines (\oi, \oii, \oiii) implemented into \spex. \\


\begin{table}[ht]
\small
\centering
\caption{\oi lines implemented into \spex}
\begin{tabular}{p{2cm}p{2cm}p{2cm}}
\hline
$ \rm \lambda (\AA)$ & E(eV) & $f_{osc}$\\
 \hline
22.5760	&	549.1673	&	$	4.70\times 10^{-6}	$	\\

22.5760	&	549.1673	&	$	1.03\times 10^{-5}	$	\\

22.5761	&	549.1648	&	$	6.31\times 10^{-6}	$	\\

22.5772	&	549.1381	&	$	6.55\times 10^{-6}	$	\\

22.5773	&	549.1356	&	$	9.08\times 10^{-6}	$	\\

22.5773	&	549.1356	&	$	1.45\times 10^{-5}	$	\\

22.5778	&	549.1235	&	$	9.27\times 10^{-6}	$	\\

22.5781	&	549.1155	&	$	1.32\times 10^{-5}	$	\\

22.5785	&	549.1075	&	$	2.08\times 10^{-5}	$	\\

22.5788	&	549.0995	&	$	1.36\times 10^{-5}	$	\\

22.5791	&	549.0915	&	$	1.95\times 10^{-5}	$	\\

22.5794	&	549.0835	&	$	3.08\times 10^{-5}	$	\\

22.5798	&	549.0755	&	$	2.07\times 10^{-5}	$	\\

22.5801	&	549.0675	&	$	4.88\times 10^{-5}	$	\\

22.5804	&	549.0596	&	$	3.08\times 10^{-5}	$	\\

22.5808	&	549.0516	&	$	2.69\times 10^{-5}	$	\\

22.5811	&	549.0436	&	$	7.50\times 10^{-5}	$	\\

22.5814	&	549.0356	&	$	4.90\times 10^{-5}	$	\\

22.5817	&	549.0276	&	$	2.60\times 10^{-6}	$	\\

22.5991	&	548.6059	&	$	3.58\times 10^{-6}	$	\\

22.6000	&	548.5841	&	$	4.58\times 10^{-5}	$	\\

22.6003	&	548.5768	&	$	1.26\times 10^{-4}	$	\\

22.6004	&	548.5744	&	$	8.36\times 10^{-5}	$	\\

22.6225	&	548.0385	&	$	8.28\times 10^{-5}	$	\\

22.6234	&	548.0167	&	$	6.55\times 10^{-6}	$	\\

22.6244	&	547.9924	&	$	8.32\times 10^{-6}	$	\\

22.6246	&	547.9876	&	$	1.18\times 10^{-4}	$	\\

22.6247	&	547.9852	&	$	3.56\times 10^{-4}	$	\\

22.6251	&	547.9755	&	$	2.59\times 10^{-4}	$	\\

22.6861	&	546.5020	&	$	4.06\times 10^{-4}	$	\\

22.6904	&	546.3985	&	$	1.05\times 10^{-4}	$	\\

22.6913	&	546.3768	&	$	1.59\times 10^{-4}	$	\\

22.6915	&	546.3720	&	$	5.88\times 10^{-5}	$	\\

22.6922	&	546.3551	&	$	4.52\times 10^{-4}	$	\\

22.6926	&	546.3455	&	$	1.22\times 10^{-3}	$	\\

22.7727	&	544.4238	&	$	4.79\times 10^{-6}	$	\\

22.7735	&	544.4047	&	$	4.42\times 10^{-6}	$	\\

22.7735	&	544.4047	&	$	1.79\times 10^{-5}	$	\\

22.7736	&	544.4023	&	$	2.62\times 10^{-6}	$	\\

22.7736	&	544.4023	&	$	2.64\times 10^{-6}	$	\\

22.7736	&	544.4023	&	$	2.53\times 10^{-6}	$	\\

22.7739	&	544.3951	&	$	4.78\times 10^{-6}	$	\\

22.7740	&	544.3927	&	$	3.24\times 10^{-6}	$	\\

22.7746	&	544.3784	&	$	8.22\times 10^{-5}	$	\\

22.7747	&	544.3760	&	$	1.93\times 10^{-5}	$	\\

22.7748	&	544.3736	&	$	1.21\times 10^{-5}	$	\\

22.7748	&	544.3736	&	$	6.07\times 10^{-6}	$	\\

22.7749	&	544.3712	&	$	8.63\times 10^{-6}	$	\\

22.7749	&	544.3712	&	$	1.69\times 10^{-5}	$	\\

22.7749	&	544.3712	&	$	9.12\times 10^{-6}	$	\\

\hline
\end{tabular}
\tablefoot{$\rm \lambda$ and E are the line wavelength and energy, respectively. $ f_{osc}$ is the oscillator strength and it is dimentionless.}
\end{table}

\begin{table}[ht]
\small
\centering
\caption{\oi lines implemented into \spex}
\begin{tabular}{p{2cm}p{2cm}p{2cm}}
\hline
$ \rm \lambda (\AA)$ & E(eV) & $f_{osc}$\\
 \hline
22.7756	&	544.3545	&	$	6.55\times 10^{-6}	$	\\

22.7758	&	544.3497	&	$	1.11\times 10^{-5}	$	\\

22.7760	&	544.3449	&	$	6.93\times 10^{-5}	$	\\

22.7761	&	544.3425	&	$	2.62\times 10^{-5}	$	\\

22.7762	&	544.3401	&	$	3.06\times 10^{-5}	$	\\

22.7765	&	544.3330	&	$	3.43\times 10^{-6}	$	\\

22.7766	&	544.3306	&	$	2.82\times 10^{-6}	$	\\

22.7766	&	544.3306	&	$	3.83\times 10^{-6}	$	\\

22.7766	&	544.3306	&	$	4.21\times 10^{-6}	$	\\

22.7777	&	544.3043	&	$	3.80\times 10^{-5}	$	\\

22.7777	&	544.3043	&	$	8.45\times 10^{-5}	$	\\

22.7779	&	544.2995	&	$	4.16\times 10^{-5}	$	\\

22.7789	&	544.2756	&	$	1.98\times 10^{-5}	$	\\

22.7790	&	544.2732	&	$	2.27\times 10^{-6}	$	\\

22.7791	&	544.2708	&	$	7.19\times 10^{-6}	$	\\

22.7792	&	544.2685	&	$	6.95\times 10^{-6}	$	\\

22.7792	&	544.2685	&	$	7.73\times 10^{-6}	$	\\

22.7801	&	544.2470	&	$	1.08\times 10^{-4}	$	\\

22.7802	&	544.2446	&	$	4.81\times 10^{-5}	$	\\

22.7804	&	544.2398	&	$	6.03\times 10^{-5}	$	\\

22.7817	&	544.2087	&	$	6.05\times 10^{-6}	$	\\

22.7822	&	544.1968	&	$	9.35\times 10^{-6}	$	\\

22.7827	&	544.1848	&	$	3.33\times 10^{-5}	$	\\

22.7828	&	544.1825	&	$	2.58\times 10^{-6}	$	\\

22.7830	&	544.1777	&	$	1.22\times 10^{-5}	$	\\

22.7831	&	544.1753	&	$	1.41\times 10^{-5}	$	\\

22.7831	&	544.1753	&	$	1.39\times 10^{-5}	$	\\

22.7839	&	544.1562	&	$	1.18\times 10^{-4}	$	\\

22.7840	&	544.1538	&	$	5.50\times 10^{-5}	$	\\

22.7843	&	544.1466	&	$	9.08\times 10^{-5}	$	\\

22.7873	&	544.0750	&	$	1.22\times 10^{-5}	$	\\

22.7880	&	544.0583	&	$	3.84\times 10^{-5}	$	\\

22.7886	&	544.0440	&	$	3.93\times 10^{-5}	$	\\

22.7889	&	544.0368	&	$	2.20\times 10^{-5}	$	\\

22.7891	&	544.0320	&	$	3.51\times 10^{-5}	$	\\

22.7891	&	544.0320	&	$	2.82\times 10^{-5}	$	\\

22.7896	&	544.0201	&	$	1.14\times 10^{-4}	$	\\

22.7899	&	544.0129	&	$	5.20\times 10^{-5}	$	\\

22.7902	&	544.0058	&	$	1.33\times 10^{-4}	$	\\

22.7976	&	543.8292	&	$	1.19\times 10^{-4}	$	\\

22.7983	&	543.8125	&	$	5.00\times 10^{-4}	$	\\

22.7998	&	543.7767	&	$	6.27\times 10^{-5}	$	\\

22.8001	&	543.7695	&	$	1.28\times 10^{-5}	$	\\

22.8001	&	543.7695	&	$	6.65\times 10^{-5}	$	\\

22.8003	&	543.7648	&	$	1.60\times 10^{-4}	$	\\

22.8006	&	543.7576	&	$	9.33\times 10^{-5}	$	\\

22.8008	&	543.7529	&	$	1.22\times 10^{-4}	$	\\

22.8013	&	543.7409	&	$	4.09\times 10^{-5}	$	\\

22.8015	&	543.7362	&	$	2.34\times 10^{-4}	$	\\

22.8193	&	543.3120	&	$	1.18\times 10^{-4}	$	\\

22.8200	&	543.2954	&	$	5.41\times 10^{-4}	$	\\

22.8233	&	543.2168	&	$	4.05\times 10^{-6}	$	\\

22.8236	&	543.2097	&	$	4.01\times 10^{-4}	$	\\

22.8238	&	543.2049	&	$	1.18\times 10^{-5}	$	\\

22.8240	&	543.2001	&	$	2.81\times 10^{-4}	$	\\

22.8249	&	543.1787	&	$	9.06\times 10^{-4}	$	\\

22.8251	&	543.1740	&	$	4.02\times 10^{-5}	$	\\

22.8258	&	543.1573	&	$	2.78\times 10^{-4}	$	\\

22.8264	&	543.1430	&	$	8.49\times 10^{-6}	$	\\

22.8275	&	543.1169	&	$	2.70\times 10^{-6}	$	\\

22.8754	&	541.9796	&	$	4.28\times 10^{-4}	$	\\

22.8761	&	541.9630	&	$	1.55\times 10^{-3}	$	\\

22.8858	&	541.7333	&	$	1.09\times 10^{-3}	$	\\

22.8882	&	541.6765	&	$	7.03\times 10^{-5}	$	\\

22.8888	&	541.6623	&	$	8.73\times 10^{-4}	$	\\

22.8898	&	541.6386	&	$	4.28\times 10^{-3}	$	\\

22.8931	&	541.5606	&	$	6.56\times 10^{-6}	$	\\

22.8937	&	541.5464	&	$	3.99\times 10^{-5}	$	\\

23.5100	&	527.3501	&	$	3.46\times 10^{-2}	$	\\

23.5114	&	527.3187	&	$	1.04\times 10^{-1}	$	\\

\hline
\end{tabular}
\end{table}


\begin{table}[ht]
\small
\centering
\caption{\oii lines implemented into \spex}
\begin{tabular}{p{2cm}p{2cm}p{2cm}}
\hline
$ \rm \lambda (\AA)$ & E(eV) & $f_{osc}$\\
 \hline
21.6791	&	571.8872	&	$	1.59\times 10^{-4}	$	\\
21.6791	&	571.8872	&	$	2.37\times 10^{-4}	$	\\
21.6992	&	571.3575	&	$	2.19\times 10^{-4}	$	\\
21.6992	&	571.3575	&	$	3.24\times 10^{-4}	$	\\
21.6993	&	571.3548	&	$	1.01\times 10^{-4}	$	\\
21.7183	&	570.8550	&	$	3.23\times 10^{-4}	$	\\
21.7184	&	570.8524	&	$	1.08\times 10^{-4}	$	\\
21.7184	&	570.8524	&	$	2.16\times 10^{-4}	$	\\
21.7353	&	570.4085	&	$	4.91\times 10^{-4}	$	\\
21.7353	&	570.4085	&	$	7.29\times 10^{-4}	$	\\
21.7354	&	570.4059	&	$	2.47\times 10^{-4}	$	\\
21.7399	&	570.2878	&	$	2.93\times 10^{-4}	$	\\
21.7400	&	570.2852	&	$	1.88\times 10^{-4}	$	\\
21.7720	&	569.4470	&	$	3.11\times 10^{-4}	$	\\
21.7722	&	569.4418	&	$	1.02\times 10^{-4}	$	\\
21.7722	&	569.4418	&	$	2.05\times 10^{-4}	$	\\
21.8077	&	568.5148	&	$	1.39\times 10^{-3}	$	\\
21.8078	&	568.5122	&	$	9.37\times 10^{-4}	$	\\
21.8079	&	568.5096	&	$	4.72\times 10^{-4}	$	\\
21.8367	&	567.7598	&	$	1.37\times 10^{-3}	$	\\
21.8368	&	567.7572	&	$	4.51\times 10^{-4}	$	\\
21.8368	&	567.7572	&	$	9.07\times 10^{-4}	$	\\
21.9018	&	566.0722	&	$	3.84\times 10^{-4}	$	\\
21.9020	&	566.0670	&	$	7.68\times 10^{-4}	$	\\
21.9022	&	566.0619	&	$	1.13\times 10^{-3}	$	\\
21.9651	&	564.4409	&	$	3.54\times 10^{-3}	$	\\
21.9653	&	564.4357	&	$	2.35\times 10^{-3}	$	\\
21.9654	&	564.4332	&	$	1.17\times 10^{-3}	$	\\
21.9829	&	563.9838	&	$	6.11\times 10^{-4}	$	\\
21.9831	&	563.9787	&	$	4.44\times 10^{-4}	$	\\
21.9832	&	563.9761	&	$	2.32\times 10^{-4}	$	\\
21.9973	&	563.6146	&	$	4.54\times 10^{-3}	$	\\
21.9974	&	563.6121	&	$	2.99\times 10^{-3}	$	\\
21.9975	&	563.6095	&	$	1.49\times 10^{-3}	$	\\
22.0086	&	563.3252	&	$	1.59\times 10^{-3}	$	\\
22.0086	&	563.3252	&	$	3.17\times 10^{-3}	$	\\
22.0086	&	563.3252	&	$	4.75\times 10^{-3}	$	\\
22.0335	&	562.6886	&	$	1.38\times 10^{-4}	$	\\
22.0335	&	562.6886	&	$	2.76\times 10^{-4}	$	\\
22.0335	&	562.6886	&	$	4.15\times 10^{-4}	$	\\
22.0961	&	561.0945	&	$	1.57\times 10^{-3}	$	\\
22.0961	&	561.0945	&	$	3.14\times 10^{-3}	$	\\
22.0962	&	561.0920	&	$	4.71\times 10^{-3}	$	\\
22.2625	&	556.9006	&	$	1.28\times 10^{-2}	$	\\
22.2625	&	556.9006	&	$	4.28\times 10^{-3}	$	\\
22.2625	&	556.9006	&	$	8.55\times 10^{-3}	$	\\
23.3499	&	530.9659	&	$	3.35\times 10^{-2}	$	\\
23.3506	&	530.9500	&	$	6.70\times 10^{-2}	$	\\
23.3517	&	530.9249	&	$	1.01\times 10^{-1}	$	\\
\hline
\end{tabular}
\tablefoot{$\rm \lambda$ and E are the line wavelength and energy, respectively. $ f_{osc}$ is the oscillator strength and it is dimensionless.}
\end{table}


\begin{table}[ht]
\small
\centering
\caption{\oiii lines implemented into \spex}
\begin{tabular}{p{2cm}p{2cm}p{2cm}}
\hline
$ \rm \lambda (\AA)$ & E(eV) & $f_{osc}$\\
 \hline
20.7883	&	596.3932	&	$	1.24\times 10^{-3}	$	\\
20.7956	&	596.1838	&	$	3.05\times 10^{-4}	$	\\
20.7975	&	596.1293	&	$	9.58\times 10^{-4}	$	\\
20.8260	&	595.3136	&	$	2.57\times 10^{-4}	$	\\
20.8286	&	595.2392	&	$	2.23\times 10^{-4}	$	\\
20.8519	&	594.5741	&	$	1.36\times 10^{-3}	$	\\
20.8625	&	594.2720	&	$	5.32\times 10^{-4}	$	\\
20.8646	&	594.2122	&	$	1.50\times 10^{-3}	$	\\
20.8905	&	593.4755	&	$	5.66\times 10^{-4}	$	\\
20.8938	&	593.3818	&	$	7.06\times 10^{-4}	$	\\
20.9735	&	591.1269	&	$	2.87\times 10^{-3}	$	\\
21.0036	&	590.2798	&	$	6.80\times 10^{-3}	$	\\
21.0101	&	590.0971	&	$	1.00\times 10^{-2}	$	\\
21.0280	&	589.5948	&	$	3.77\times 10^{-4}	$	\\
21.0284	&	589.5836	&	$	2.63\times 10^{-4}	$	\\
21.0322	&	589.4771	&	$	1.73\times 10^{-4}	$	\\
21.0784	&	588.1851	&	$	3.28\times 10^{-3}	$	\\
21.0859	&	587.9759	&	$	6.38\times 10^{-3}	$	\\
21.0945	&	587.7361	&	$	1.21\times 10^{-3}	$	\\
21.0957	&	587.7027	&	$	1.71\times 10^{-4}	$	\\
21.1666	&	585.7341	&	$	1.11\times 10^{-3}	$	\\
21.2056	&	584.6569	&	$	4.75\times 10^{-4}	$	\\
21.2066	&	584.6293	&	$	1.35\times 10^{-2}	$	\\
21.2240	&	584.1500	&	$	4.84\times 10^{-3}	$	\\
21.2855	&	582.4622	&	$	2.23\times 10^{-3}	$	\\
21.3140	&	581.6834	&	$	2.44\times 10^{-3}	$	\\
21.3252	&	581.3779	&	$	4.79\times 10^{-3}	$	\\
21.3428	&	580.8985	&	$	3.40\times 10^{-3}	$	\\
21.3584	&	580.4742	&	$	6.00\times 10^{-3}	$	\\
21.5092	&	576.4045	&	$	1.33\times 10^{-2}	$	\\
21.5313	&	575.8129	&	$	2.34\times 10^{-2}	$	\\
21.5836	&	574.4176	&	$	5.45\times 10^{-3}	$	\\
22.9400	&	540.4534	&	$	6.89\times 10^{-2}	$	\\
23.0280	&	538.3880	&	$	1.04\times 10^{-1}	$	\\
23.0710	&	537.3846	&	$	1.26\times 10^{-1}	$	\\
\hline
\end{tabular}
\end{table}
\end{document}